\documentclass[twocolumn,showpacs,nofootinbib,aps,prd,amsmath,amssymb,nobibnotes,floatfix]{revtex4-1}

\usepackage{hyperref}
\usepackage{amsfonts}
\usepackage{mathrsfs}
\usepackage{epsfig}
\usepackage{graphicx}               
\usepackage{url}
\usepackage{hyperref}
\usepackage{float}
\usepackage{color}

\usepackage{graphicx}
\usepackage{epsf}
\usepackage[utf8]{inputenc}

\usepackage{comment}

\newcommand{\nc}{\newcommand}

\nc{\beq}{\begin{equation}}
\nc{\eeq}{\end{equation}}
\nc{\beqa}{\begin{eqnarray}}
\nc{\eeqa}{\end{eqnarray}}

\usepackage{slashed}

\newcommand{\lsim}{\!\mathrel{\hbox{\rlap{\lower.55ex \hbox{$\sim$}} \kern-.34em \raise.4ex \hbox{$<$}}}}
\newcommand{\gsim}{\!\mathrel{\hbox{\rlap{\lower.55ex \hbox{$\sim$}} \kern-.34em \raise.4ex \hbox{$>$}}}}

\newcommand{\vev}[1]{ \left\langle {#1} \right\rangle }

\def\be{\begin{equation}}
\def\ee{\end{equation}}

\def\dps{\bar{\delta}_{\rm PS}}

\newcommand\affspc{\vspace{4pt}}

\usepackage{footmisc}
\usepackage{setspace}

\begin{document}

\title{Comparing fully general relativistic and Newtonian calculations of structure formation}

\author{William E.\ East$^1$, Rados\l{}aw Wojtak$^{2,3}$, and Tom Abel$^2$}
\affiliation{$^1$Perimeter Institute for Theoretical Physics, Waterloo, Ontario N2L 2Y5, Canada \affspc}
\affiliation{$^2$Kavli Institute for Particle Astrophysics and Cosmology, Stanford
University, SLAC National Accelerator Laboratory, Menlo Park, California 94025, USA \affspc}
\affiliation{$^3$Dark Cosmology Centre, Niels Bohr Institute, University of
  Copenhagen, Juliane Maries Vej 30, \\ DK-2100 Copenhagen \O,
  Denmark\affspc}

\begin{abstract}
In the standard approach to studying cosmological structure formation, the
overall expansion of the Universe is assumed to be homogeneous, with the
gravitational effect of inhomogeneities encoded entirely in a Newtonian
potential. A topic of ongoing debate is to what degree this fully captures the
dynamics dictated by general relativity, especially in the era of precision
cosmology.  To quantitatively assess this, we directly compare standard N-body
Newtonian calculations to full numerical solutions of the Einstein equations,
for cold matter with various magnitude initial inhomogeneities on scales
comparable to the Hubble horizon.  We analyze the differences 
in the evolution of density, luminosity
distance, and other quantities defined with respect to fiducial observers.
This is carried out by reconstructing the effective spacetime and matter fields
dictated by the Newtonian quantities, and by taking care to distinguish effects
of numerical resolution.  We find that the fully general relativistic and
Newtonian calculations show excellent agreement, even well into the nonlinear
regime. They only notably differ in regions where
the weak gravity assumption breaks down, which arise when considering extreme cases 
with perturbations exceeding standard values.
\end{abstract}

\maketitle

\section{Introduction}
Our observable Universe appears to be, to a good approximation, homogeneous,
isotropic, and flat on the largest scales, but with rich structure at smaller
scales. The usual approach in cosmology is to treat the Universe on large
scales as governed by a homogeneous solution to the Einstein equations---the
Friedmann-Roberson-Walker (FRW) solution---with small deviations away from
homogeneity, which are treated perturbatively. Smaller scales, where deviations from
homogeneity become large and lead to formation of clusters of galaxies and
other structures, are assumed to be decoupled from the large-scale dynamics. 
However, this treatment is only an approximation, as
general relativity (GR) is inherently a nonlinear system which couples
different scales.  Recently, there has been much interest in applying advances
in numerically solving the full Einstein equations to study inhomogeneous
cosmologies~\cite{Bentivegna:2015flc,Giblin:2015vwq,Giblin:2016mjp,Macpherson:2016ict,Daverio:2016hqi}.
Such studies are motivated by assessing ``backreaction" effects---that is
the potential for smaller-scale inhomogeneities to effect the overall expansion
of the Universe, a topic that remains
controversial~\cite{Buchert:1999er,Kolb:2004am,Rasanen:2011ki,Ishibashi:2005sj,Green:2014aga}
---and, in general, quantifying relativistic effects and their possible impact
on making measurements in the era of precision cosmology.
Impetus for this is provided by ongoing and upcoming cosmological surveys such
as the Dark Energy Survey (DES) (see year-one results in~\cite{des1yr2017}),
the Dark Energy Spectroscopic Instrument (DESI) \cite{desi2013}, the Large
Synoptic Survey Telescope (LSST) \cite{lsst2012}, or the Euclid space mission
\cite{euclid2011}. These surveys will provide vast observational data, from cosmological
distances to lensing observations, measured with unprecedented precision, for
testing assumptions underlying the standard cosmological model.

Studies utilizing full GR solutions have begun to explore the nonlinear effects
that appear for sufficiently large inhomogeneities.  However, standard
Newtonian cosmology simulations also capture effects in collapse and structure
formation that are nonlinear in the amplitude of the inhomogeneities, so the
important question that remains to be answered is, how important are effects
that are both nonlinear and relativistic? The goal of this work is to realize a
meaningful comparison between standard Newtonian cosmology calculations, and
those utilizing full GR, that will allow us to quantify how much the two types
differ. 

To do this, we directly compare Newtonian and full GR cosmological simulations
of cold matter in an expanding Universe. For the former, we use standard N-body
techniques to evolve a set of particles on a FRW background that source
(through a Poisson-type equation), and respond to, a Newtonian gravitational
potential. For the latter, we numerically solve the nonlinear constraint (at
the initial time) and evolution parts of the Einstein equations, using standard
grid-based methods.  Given the computational expense of solving the full
Einstein field equations, instead of using entirely realistic initial
conditions, we focus on some simplified setups that contain inhomogeneities at
a modest range of scales and allow us to compare the two types of calculations
as a function of their amplitude.  In this work we study a range of cosmological
models, including ones where the density fluctuations exceed the rms 
of the density field at the corresponding scales in the standard $\Lambda$CDM
cosmological model by factor of up to $\sim100$ (i.e., they roughly correspond to
rms $\sim 0.1$ at a Gpc scale at the present time). This is partly considered as a
limiting case, to see how large the amplitude of the inhomogeneities can be
made before significant relativistic effects arise.  However, we also note that
the possibility that high over(under)density structures are present on scales
larger than the baryon acoustic oscillations have not been fully ruled out. For
example, several studies find evidence for a $\sim300$ Mpc underdensity in the
southern sky, detected both in the distribution of galaxies \citep{Keenan2013}
and x-ray galaxy clusters \citep{Boe2015}.

We carry out the comparison in terms of quantities defined with respect to a 
set of fiducial observers, e.g.  luminosity distance-redshift relations, both
since these are the most readily interpreted and relevant quantities, and
because this will obviate difficulties associated with the different
coordinates used in the Newtonian and GR calculations. In order to facilitate
this comparison, and to ensure that we are setting up equivalent initial
conditions in the two cases, we make use of a dictionary that allows an
effective spacetime and set of matter fields to be reconstructed from the
evolution variables of the Newtonian
simulation~\cite{Chisari:2011iq,Green:2011wc} (once the density fields have
been suitably constructed from the particles using the techniques
of~\cite{Abel2012}).  We find that the Newtonian calculations, suitably
interpreted, in fact agree quite well with the full GR results, well into the
nonlinear regime.  We only find a significant difference in extreme cases where
the magnitude of the Newtonian potential is no longer much smaller than 1.
We also comment on the possible differences that can arise
due to using a fluid versus particle description for matter,
as is commonly done in conjunction with the different approaches
to gravity.

Previous studies utilizing evolutions in full GR have mainly focused on
comparing to linear theory using simple setups with perturbations initially at
a single length scale, and following the evolution of matter or certain
metric functions~\cite{Bentivegna:2015flc,Macpherson:2016ict}, or evolved
perturbations at a range of length scales over an increase in scale factor 
by a factor of a few, while
also tracking light propagation~\cite{Giblin:2015vwq,Giblin:2016mjp}.  In this
work we consider initial inhomogeneities of both these types,
evolved through a $\sim100\times$ increase in scale factor.  We also use
initial data that nontrivially solves the momentum constraint of the Einstein
equations. This contrasts with previous treatments that trivially satisfy the
momentum constraint by assuming a moment of time symmetry---and hence include
decaying, as well as growing, perturbations---or do not solve the nonlinear
momentum constraint. In addition, as in~\cite{Macpherson:2016ict}, we use a
treatment that is not restricted to synchronous gauge (geodesic slicing), where the lapse is set
to unity and the shift vector to zero, which will break down when caustics form.\footnote{Geodesic
slicing is also not strongly hyperbolic in the Baumgarte-Shapiro-Shibata-Nakamura formulation of
the Einstein equations~\cite{Beyer:2004sv}.}
This comes at the expense of having to also keep track of the nontrivial
evolution of matter.

Tackling the problem from the other end, there has also been work 
comparing N-body calculations to exact solutions of the Einstein
equations~\cite{Alonso:2010zv}, and 
incorporating
various relativistic effects into such calculations, for example evolving
additional metric degrees of freedom in the weak gravity
limit~\cite{Adamek:2015eda,Adamek:2017uiq}, or including relativistic screening
through a Helmholtz equation~\cite{Hahn2016}.  In addition, as a way to probe
the behavior of inhomogeneities on cosmic expansion in the extreme relativistic
limit, there have been studies using full GR solutions of black hole
lattices~\cite{Bentivegna:2012ei,Yoo:2013yea,Yoo:2014boa}.  

The rest of this paper is organized as follows.  In Sec.~\ref{sec:dict} we
review the relativistic-Newtonian matching scheme that we will use in setting
up equivalent initial conditions and making comparisons.  In
Sec.~\ref{sec:method}, we describe the initial conditions for the various cases
we consider, outline how we perform the respective Newtonian and GR
calculations, and describe how we define and compute various ``observable"
quantities that we will compare between the two cases.  The results of this
comparison are given in Sec.~\ref{sec:results}.  We conclude in
Sec.~\ref{sec:conclusion} and mention some directions for future work.  In the
Appendix we describe results from resolution studies used to assess numerical
error. Unless otherwise stated, we use units with $G=c=1$ throughout.

\section{Relativistic translation of Newtonian quantities}
\label{sec:dict}
In this paper, we consider solutions of general relativity coupled to a matter
model consisting of pressureless fluid in a periodic domain, and
compare this to the N-body simulations of Newtonian gravity on the background of an
expanding FRW solution commonly used in studies of structure formation.
Properly interpreted, the quantities from such simulations should agree both
with linear perturbation theory for sufficiently small perturbations around a
homogeneous FRW solution, and with nonlinear Newtonian gravity on scales much
smaller than the Hubble radius.  In \cite{Chisari:2011iq,Green:2011wc} a
relativistic-Newtonian matching scheme is laid out that we will use to set up
equivalent initial conditions and compare quantities between the GR and
Newtonian calculations.  In this section we briefly review this scheme.

For the Newtonian simulations we assume a background FRW solution with density
$\rho_0$, scale factor $a$, and Hubble parameter $H$.  We then calculate, on the
simulation domain, a density $\rho_N$, gravitational potential $\psi_N$, and a
velocity $v^i$.  From the density we can also define a density contrast
$\delta_N$:
\beq  
\rho_N = \rho_0(1+\delta_N).
\eeq
The gravitational potential $\psi_N$ satisfies
\beq  
\partial^i \partial_i \psi_N = 4 \pi a^2 \rho_0 \delta_N
\label{eq:poisson}
\eeq
and the evolution of the density perturbation
\beq
\dot{\delta}_N+\partial_i ((1+\delta_N)v^i) = 0
\label{eq:ddot}
\eeq
where the derivatives are with respect to comoving coordinates and conformal time $\tau$.

Under some simplifying assumptions the metric that we can reconstruct from the Newtonian quantities is:
\beq
ds^2 = a^2\left[-(1+2\psi_N)d\tau^2+(1-2\psi_N)\delta_{ij}dx^i dx^j\right].
\label{eq:newt_metric}
\eeq 

The quantities that make up the stress-energy tensor $T^{ab}=\rho u^a u^b$
in the relativistic treatment are as follows.
The density is given by
\beq
\rho/\rho_0-1=\delta = \delta_N -2\psi_N -2 d\psi_N/d \log(a)
\label{eq:dens_dict}
\eeq
and the four-velocity is 
\beq
u^a=u^{\tau}(1,v^i)
\eeq
where the time component can be calculated from the normalization requirement $u^au_a=-1$ as 
\beq
(u^{\tau})^{-1} = a\sqrt{1+2\psi_N-(1-2\psi_N)\delta_{ij}v^iv^j}. 
\eeq
We also note that along the trajectory of some observer or particle, we can calculate the proper time
as 
$t_p = \int (u^{\tau})^{-1} d \tau$.

In the above, we have ignored the vector modes of the metric, both because they
are expected to be small, and because determining them would require the
solution of additional elliptic equations that are not typically solved in
Newtonian simulations (though see~\cite{Bruni:2013mua,Thomas:2015kua}).  This is the correspondence in~\cite{Chisari:2011iq}, and
in the ``abridged dictionary" of~\cite{Green:2011wc}.  The goal of this work
will be to quantify how closely the spacetime metric and matter fields
constructed from the Newtonian solution above match the full solution of the
Einstein equations.

\section{Methodology}
\label{sec:method}
\subsection{Initial conditions}
In this section we detail the initial conditions we use. We begin with the initial data for the Newtonian
simulations. We then outline how these translate into the GR quantities and specify how we solve the constraint
part of the Einstein equations to obtain fully relativistic initial data for the GR calculations. 
For convenience we will assume that
at the initial time $a=1$.

For the Newtonian simulations, we specify the density perturbations and velocities $\{\delta_N,v^i\}$.
We take the density perturbations to be a sum over modes with 
different amplitudes $\bar{\delta}_n$, wave numbers ${\bf k}_n$, and phases $\phi_n$:
\beq
\delta_N = \sum_n \bar{\delta}_n \sin({\bf k}_n\cdot {\bf x}+\phi_n).
\label{eq:gen_id}
\eeq
For many of the cases we consider, we will use a simplified
version of this where all of the components of the density perturbations have the
same amplitude and wave number magnitude in each of the coordinate directions:
\beq
\delta_N = \bar{\delta} \sum_i \sin(k x^i) .
\label{eq:simple_id}
\eeq

For the velocity initial condition, we use the Zel'dovich approximation (ZA) \cite{Zeldovich1970}:
\beq
{\bf v}= H \sum_n {\bf k}_n\bar{\delta}_n \cos({\bf k}_n\cdot {\bf x}+\phi_n)/k^2_n. 
\label{eq:v_zd}
\eeq
For comparison with previous work where initial data were chosen to trivially
satisfy the momentum constraint (e.g.~\cite{Bentivegna:2015flc,Giblin:2015vwq}), we also consider a case where the velocity is initially
zero: $v^i=0$.

Once we have specified $\{\delta_N,v^i\}$, 
we can calculate $\psi_N$ from Eq.~(\ref{eq:poisson}),
and thus $\dot{\psi}_N$, in order to 
calculate the relativistic quantities.
Taking the time derivative of Eq.~(\ref{eq:poisson}), and combining it with Eq.~(\ref{eq:ddot}) (dropping
the second-order term), we have that 
\beq
\partial^i\partial_i \dot{\psi}_N \approx 
-\frac{3}{2}H^2(\partial_i v^i+H\delta_N) . 
\eeq
This can be inverted to give an approximation of $\dot{\psi}_N$
at the initial time.
For the Zel'dovich approximation velocity profile this just gives $\dot{\psi}_N=0$
and implies that the density perturbation is evolving with the Hubble flow:
$\dot{\delta}_N = H \delta_N$.
With these quantities in hand we can apply the dictionary of Sec.~\ref{sec:dict} to
calculate everything else.  For example, for the simple density profile 
of Eq.~(\ref{eq:simple_id}) from Eq.~(\ref{eq:dens_dict}), we have that
$\delta = \left[1+3(H/k)^2\right] \delta_N$ and $\delta=\delta_N$ for the Zel'dovich velocity profile
and the zero velocity profile, respectively. 
 
In addition to the density, the rest of the quantities for the GR initial data
can be calculated from the metric in Eq.~(\ref{eq:newt_metric}).  Note, however, that the
Einstein equations also impose constraints---the Hamiltonian and momentum
constraints---on the initial metric.  We solve these constraints in the
conformal thin-sandwich formalism using the code described
in~\cite{idsolve_paper}.  In this formalism we specify the conformal
three-metric $\tilde{\gamma}_{ij}$, the trace of the extrinsic curvature $K$,
the conformal lapse $\tilde{\alpha}$, the matter density $\rho$ and 
the conformal three-momentum $\tilde{p}^i$: 
\begin{equation}
\begin{aligned}
\tilde{\gamma}_{ij} &= (1-2\psi_N)\delta_{ij},\\
K &= -3H(1-\psi_N-\dot{\psi}_N/H),\\
\tilde{\alpha} &= \sqrt{1+2\psi_N},\\
\tilde{p}^i &= \rho_0(1+\delta) \alpha (u^{\tau})^2 v^i,
\end{aligned}
\end{equation}
as well as the
traceless part of the time derivative of the metric, which we set to zero
$\partial_t \tilde{\gamma}^{ij}=0$. 
With these free data, we solve the conformal thin-sandwich equations\footnote{
In contrast to~\cite{idsolve_paper}, we do not conformally rescale the energy.
} 
for a conformal factor $\Psi$ and shift vector 
$\beta^i$, such that the four-metric  
\beq
g_{ab} dx^a dx^b = -\Psi^{12}\tilde{\alpha}^2 dt^2+\Psi^4\tilde{\gamma}_{ij}(\beta^idt+dx^i)(\beta^j dt+ dx^j)
\eeq
satisfies the nonlinear constraint equations. 
Since the conformal quantities already satisfy the constraint equations to
linear order, we expect the quantities $(\Psi-1)$ and $\beta^i$ to be small and
to scale like $\bar{\delta}^2$ for small initial inhomogeneities, which is true
for all the cases considered here.
We will consider initial conditions that consist of small perturbations on
superhorizon scales, and our method for constructing initial conditions is in
keeping with the assumption for the validity of the Newtonian approximation
that this regime should be well described by linear perturbation theory.

\subsection{Newtonian simulations}
We carry out N-body simulations using the \emph{GADGET-2} code \cite{Springel2005} in a mode 
for following the evolution of collisionless matter. The code combines two methods to compute 
gravitational forces: the Fourier technique for the contribution from long-range forces and the 
hierarchical tree method for short-range forces. The positions and velocities of particles are 
advanced using leapfrog integration with an adjustable time step. The Newtonian evolution 
of the particles is decoupled from the background expansion which is governed by the Friedmann 
equation.

We generate initial conditions by displacing particles from the positions given by regular mesh. 
The displacement field is related to the gradient of the initial potential $\psi_{N}(a=1)$ through 
the Zel'dovich approximation \citep{Zeldovich1970}
\beq
\delta x_{i} =-\frac{1}{4\pi\rho_0}\partial_i\psi_{N}(a=1),
\eeq
where $\partial_{i}$ is the derivative with respect to the initial
Lagrangian coordinates.
Due to the nonlinearity of the transformation between the Lagrangian and Eulerian coordinates, 
the density field generated by the above displacement can slightly differ from the assumed 
initial density. The relative deviations from the analytic model given by Eq.~(\ref{eq:gen_id}) 
reach the percent level for initial conditions with the highest amplitude $\bar{\delta}$. 
In order to mitigate this problem, we alter particle masses in 
a way that they compensate differences between the actual and assumed 
density field. This correction makes the density field computed from the particle position 
resemble the analytic model with relative errors in the density contrast $\delta_{\rm N}$ 
of $10^{-3}$.

The N-body code does not explicitly evolve the density field, which needs to be computed 
from the particle positions in a postprocessing analysis. We employ a method 
based on tracing the evolution of the initial (Lagrangian) tessellation of the dark matter 
manifold in phase space \citep{Abel2012,Shandarin2012}. 
The local density is primarily determined by the expansion or contraction (in regions with no shell crossing, e.g. voids) 
and superposition (in multistream regions, e.g. halos) of tetrahedral volume elements defined by fixed groups of particles 
(neighboring particles in the initial Lagrangian space). Assuming that every particle contributes equally 
to the mass elements carried by all adjacent tetrahedra leads to a straightforward means of estimating 
the density at particle positions. Additional assumptions regarding interpolation schemes are required 
for estimating the density at arbitrary points. Here we follow the approach outlined in \citep{Abel2012}. 

The accuracy of the adopted density estimator has some limitations. Less accurate density estimates 
can be expected in multistream regions (e.g. halos) where the density estimator does 
not fully comply with the effective density of the Poisson solver in the N-body code. However, as we shall see, 
the detailed properties of the matter distribution in these regions are quite sensitive to numerical resolution 
both in the GR and N-body simulations.

\subsection{GR simulations}
To evolve the GR-hydrodynamic equations we use the code described
in~\cite{code_paper}.  The Einstein equations are evolved in a periodic domain
in the generalized harmonic formulation using a damped harmonic
gauge~\cite{Choptuik:2009ww,Lindblom:2009tu} in a similar manner as
in~\cite{East:2015ggf}.  We make our initial conditions compatible with this
choice of gauge by appropriately choosing $\partial_t g_{ta}$ (or equivalently,
the time derivatives of the lapse and shift) at the initial time, so it does
not affect the correspondence with the Newtonian quantities on the initial time
slice.  We use fourth-order Runge-Kutta time stepping and standard fourth-order
finite differences for the spatial derivatives.  

We note that stably evolving the Einstein equations requires resolving the
light-crossing time between grid cells since this is the speed at which
information propagates.  This is in contrast to Newtonian simulations, where
gravity is encapsulated in an elliptic equation, and the necessary time
resolution is set by the velocity of the particles.  This is the primary reason
that solutions of full GR are much more computationally expensive than their Newtonian
counterparts.  To deal with the fact that the metric functions grow due to
expansion, placing stricter limits on the time-step size for numerical
stability,\footnote{In particular, with the gauge choice used here, the lapse grows.} we decrease $dt$ in proportion to the minimum of $\alpha^{-1}$ over
the whole domain during evolution.

Unlike some other approaches, we have not chosen a synchronous gauge, which
means that we do not have to worry about the potential for coordinate problems
from the formation of caustics, and we can use a gauge that has been found
to be robust in the strong field, dynamical regime.  However, it does mean that the dust velocity
will not be zero in these coordinates, and the evolution of the dust will have
to be kept track of as well.  The way we handle this is just to evolve the
hydrodynamic equations 
but with a fixed, negligibly small
pressure ($P\sim10^{-12}\rho$) and to ignore the energy evolution equation.
The fluid equations are evolved as in~\cite{code_paper} using standard high-resolution shock-capturing
techniques that are second-order accurate for smooth flows and reduce to
first order in the presence of shocks.
We present details on convergence and estimates of numerical error in the Appendix.

\subsection{Particle versus fluid differences}
Since we use a particle description of matter for the Newtonian
calculations  and a fluid description for the GR calculations, there will, in
principle, be differences between the two, irrespective of their treatment of
gravity. In the particle case we are approximating the collisionless Boltzmann
equation.  Taking moments of this equation, the evolution of the density will
obey the continuity equation, while the momentum density will obey the Jeans
equation. These can be thought of as equivalent to the Euler equations
governing a fluid, but with an anisotropic effective pressure that is nonzero in
multistream regions and is set by the velocity dispersion. On the other hand, when
actually evolving a fluid in the GR case, we take the pressure to be zero.
In order to quantify this, we measure the velocity dispersion
in the N-body calculation
\beq
\sigma_v =   \vev{|\mathbf{v} - \vev{\mathbf{v}}|^2}^{1/2}, 
\eeq 
where $\vev{\ldots}$ represent an average over momentum space,
for some representative cases below.
In practice, we find these differences to be negligible for most of the
comparisons we make in this work, where the velocity dispersion is zero (in
single-stream regions) or small, and to only be significant in the vicinity of
large collapsing regions at late times. 

\subsection{Calculating observables}
To make a meaningful comparison between the Newtonian and GR calculations, we
want to utilize observable quantities---that is, quantities defined in
terms of a set of fiducial observers. This is especially important since we use
different coordinates for the two calculations.  To facilitate this we will
make use of a set of geodesics, both timelike and null, that are defined with
respect to the initial time slice where the two calculations \emph{do} make use of the
same gauge (up to small nonlinear corrections). 

One quantity we will compare is the density $\rho$ measured as a function of
proper time, as seen by a chosen set of observers comoving with matter (at
the underdensities, overdensities, etc.).  In the GR simulations this is 
calculated by integrating geodesics and evaluating $\rho$ along their
worldlines.  In the Newtonian simulations this can be calculated by saving
$\{\delta_N, v^i, \psi_N, \dot{\psi}_N\}$ along different particle trajectories
and using the formulas in Sec.~\ref{sec:dict}.  Even though the ``observer"
quantities are proper time and density, for convenience we can translate this
into an effective scale factor and density contrast by making reference to the
FRW solution (but not referring to any global or averaged quantities).  The
scale factor that a fiducial observer would get by integrating the FRW solution
as a function of proper time is just 
\beq
a_p:=\left[3t_pH(a=1)/2+1\right]^{2/3}. 
\eeq
Likewise the density is 
$\rho_{\rm FRW}(t_p)=\rho_0(a=1)a_p^{-3}$, from which we can define a density contrast from
only the observer's local quantities as 
\beq
\delta_{\rm obs}(t_p):=\rho(t_p)/\rho_{\rm FRW}(t_p)-1.
\eeq
We emphasize that this is just a convenient parametrization of the density seen by an observer
comoving with matter and will differ from the quantity $\delta_N$.

We also calculate null geodesics as a point of comparison, by directly integrating
the geodesic equation 
\beq
\frac{d k^{a}}{d \lambda}+\Gamma^{a}_{bc}k^bk^c=0
\eeq
where $k^a$ is the four-velocity, $\lambda$ is an affine parameter, and
$\Gamma^{a}_{bc}$ is the Christoffel symbol.  For the Newtonian simulations we
also directly integrate the geodesic using the values from the
reconstructed metric [Eq.~(\ref{eq:newt_metric})] as a postprocessing step. This
will, in some sense, include ``relativistic" effects in the propagation of light,
but the viewpoint we are taking is that we want to compare how similar the
spacetime given by Eq.~(\ref{eq:newt_metric}) is to the spacetime that comes from
solving the Einstein equations, and tracing out geodesics is simply a way to
measure this.

From the four-velocity of each of these null geodesics, we can compute a redshift with respect to an emitter and observer
comoving with matter 
\beq
z = -1+\frac{(u_a k^a)_{\rm emit}}{(u_a k^a)_{\rm obs}}.
\eeq
Following~\cite{Bentivegna:2016fls}, for each primary null geodesic, we also compute two neighboring null geodesics that are perturbed slightly in the directions
orthogonal to the geodesic's initial four-velocity, and 
we calculate the luminosity distance
$D_L$ (or equivalently, the angular distance, as the two quantities are related by the reciprocity relation~\cite{Etherington2007}) 
from its relation to the geodesic deviation equation. 
See~\cite{2004LRR.....7....9P,Bentivegna:2016fls} for details. 

Below we shall primarily concentrate on comparisons between the Newtonian and
fully general-relativistic calculations of $\{a_p(t_p),\delta_{\rm obs}(t_p)\}$
along specified timelike geodesics and $\{z(\lambda),D_L(\lambda)\}$ along
null geodesics.

\subsection{Cases}
\label{sec:cases}
In this study we compare the general-relativistic and Newtonian evolution
beginning from several different initial conditions, for an inhomogeneous,
dust-filled, expanding universe with vanishing global curvature (the Einstein--de
Sitter model, i.e. $\Omega=1$).  We consider several cases where the inhomogeneities are
initially at one length scale [Eq.~(\ref{eq:simple_id})], and the velocity is
given by the Zel'dovich approximation [Eq.~(\ref{eq:v_zd})].  We fix $k=\pi
H(a=1)/2$---i.e. the initial wavelength is 4 times the initial Hubble
radius---and consider various magnitudes for the inhomogeneities:
$\bar{\delta}=5\times10^{-4}$, $10^{-3}$, and $10^{-2}$.  For comparison with
previous work, we also consider initial conditions equivalent to the
$\bar{\delta}=10^{-3}$ case but with initial velocity that is identically
zero.

In addition, we consider cases with inhomogeneities at a range of length
scales.  In particular, we consider a spectrum of inhomogeneities given by
Eq.~(\ref{eq:gen_id}) where $\bar{\delta}_n$ is nonzero for $\pi/2 \leq
k_n/H(a=1) \leq 6 \pi$, and given by drawing from a normal distribution with
zero mean and $\sigma_n=(k_n/k_{\rm min})^{-3/2}$, $\bar{\delta}_n\sim
\dps\times \mathcal{N}(0,\sigma_n^2)$ with $\dps=10^{-3}$ and
$10^{-2}$.  We also chose $\phi_n$ in Eq.~(\ref{eq:gen_id}) from a uniform
distribution on $[0,2\pi]$, and again use the Zel'dovich approximation for the
initial velocity profile.

For several cases we perform simulations with multiple resolutions in order to
estimate numerical errors, which we discuss in detail in the Appendix.
Unless otherwise stated, results in the following are shown from the highest
resolution runs, utilizing $256^3$ particles for the Newtonian simulations, and
between $192^3$ and $256^3$ grid cells for the GR simulations.

\section{Results}
\label{sec:results}

\subsection{Single-wavelength initial conditions}
\label{subsec:simple_id}
We first focus on simpler initial conditions of the form given by Eq.~(\ref{eq:simple_id}),
where the inhomogeneities are
initially all at a wavelength that is 4 times the initial Hubble radius, and
follow the evolution of these inhomogeneities as they enter the horizon and
grow.  To begin with we compare a case where the initial velocity profile is
zero to one where the velocity is given by the Zel'dovich approximation. In
the latter case $|\delta_N|$ grows linearly with the scale factor in the Newtonian
picture, beginning at the initial time.  The zero-velocity initial data, on the
other hand, includes both growing and decaying density perturbations, and so
$|\delta_N|$ initially grows slower.  

\begin{figure}
\begin{center}
\includegraphics[width=\columnwidth,draft=false]{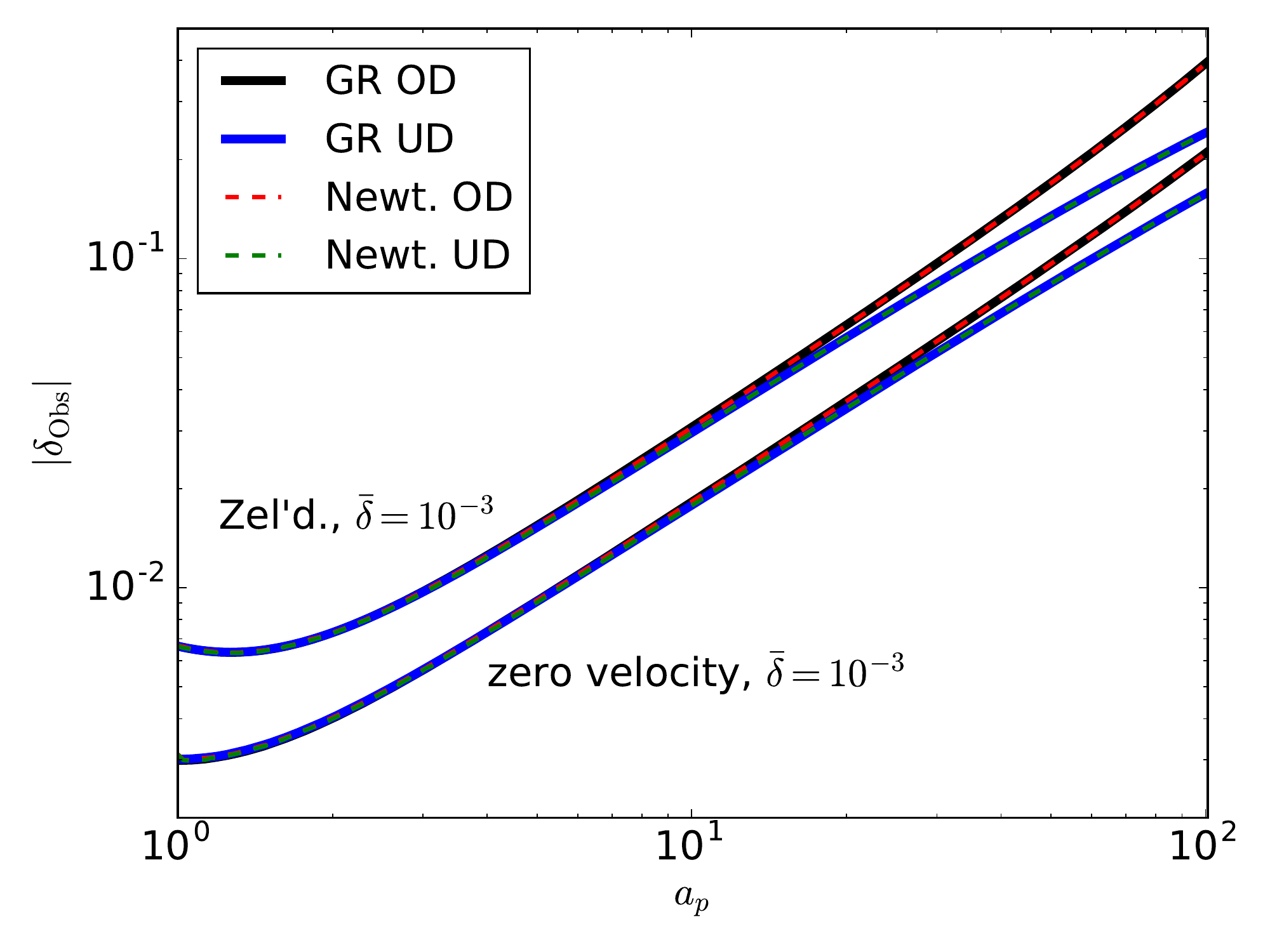}
\end{center}
\caption{
The density contrast as measured by an observer comoving with matter at the
point of maximum overdensity (black and red curves) or underdensity (blue and green
    curves) for two cases with $\bar{\delta}=10^{-3}$ in Eq.~(\ref{eq:simple_id}), and
an initial velocity profile that is either zero or given by the Zel'dovich
approximation. The general-relativistic (solid lines) and Newtonian (dotted
lines) calculations closely track each other.
\label{fig:zv_zd}
}
\end{figure}

We show the density measured by some fiducial observers for these two cases in
Fig.~\ref{fig:zv_zd}.  Though the sizes of the initial \emph{Newtonian} density
perturbations $\delta_N$ are the same in both cases [given by
Eq.~(\ref{eq:simple_id}) with $\bar{\delta}=10^{-3}$], they correspond to
different densities through the relation given by Eq.~(\ref{eq:dens_dict}).
However, making use of this correspondence between the Newtonian and GR
calculations, as illustrated in Fig.~\ref{fig:zv_zd}, both give fully consistent 
results, even as the perturbations become nonlinear---as evidenced by the
diverging of the magnitude of the density contrast at the overdensities and
underdensities.  In what follows we will focus on initial conditions given by
the Zel'dovich approximation velocity profile since this gives only growing modes,
and we will study how close the relativistic and Newtonian calculations are in the
nonlinear regime.  

\begin{figure}
\begin{center}
\includegraphics[width=\columnwidth,draft=false]{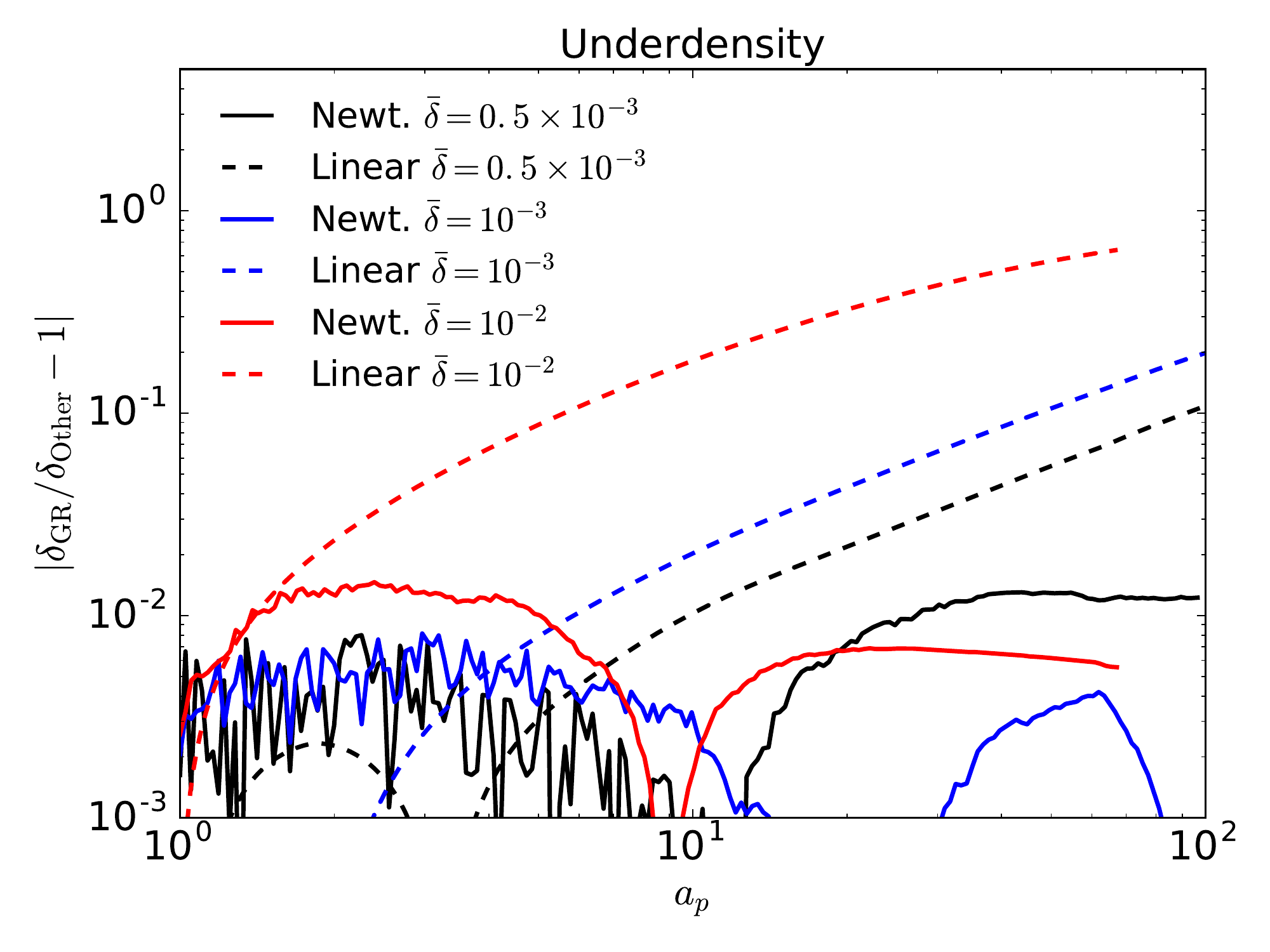}
\includegraphics[width=\columnwidth,draft=false]{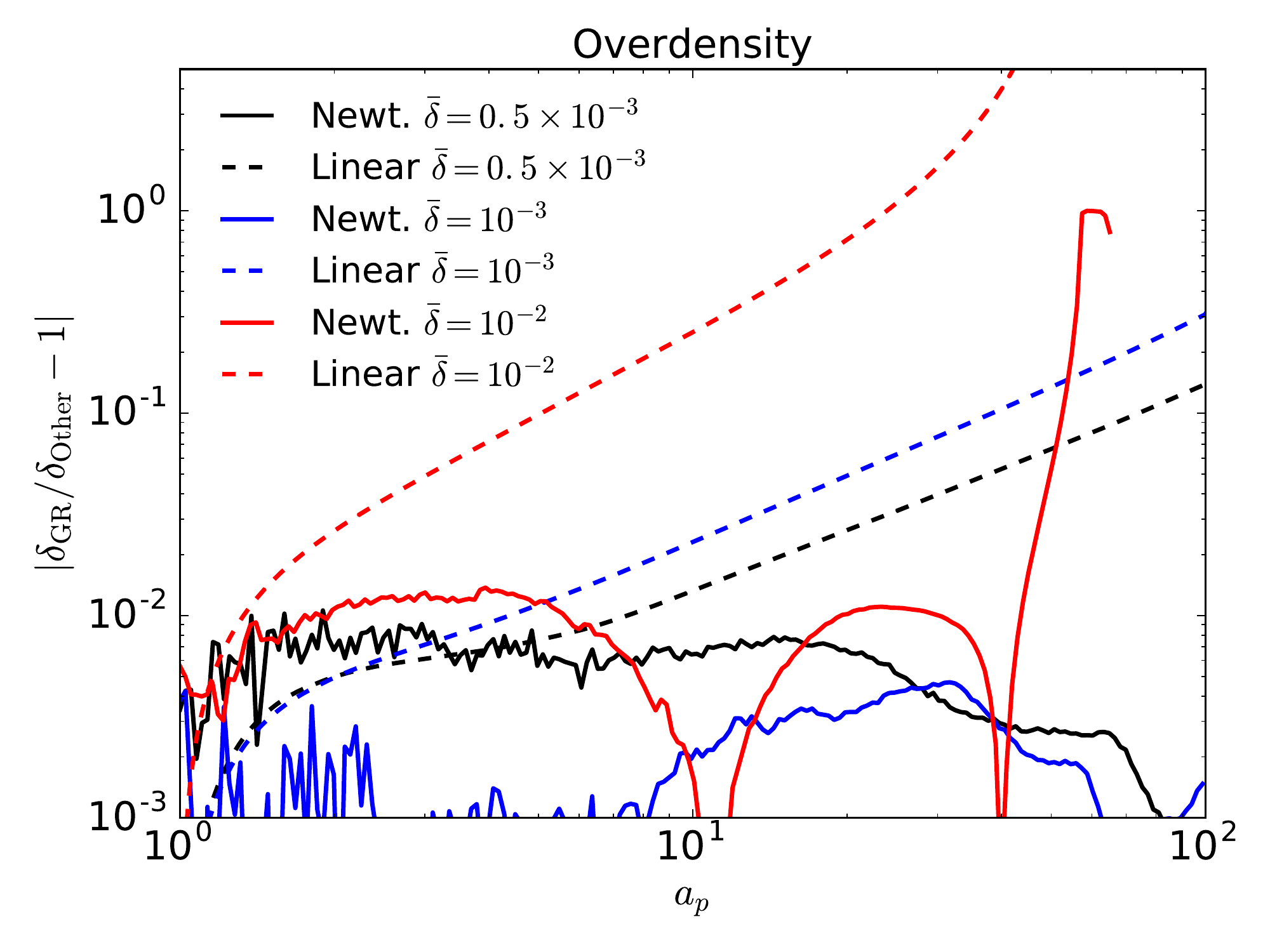}
\end{center}
\caption{
The fractional difference between the density contrast $\delta_{\rm obs}$
at an underdensity (top) or overdensity (bottom)
in the GR calculation compared to Newtonian simulations (solid lines) or 
linear perturbation theory (dotted lines) for various magnitudes of the initial
inhomogeneities.
\label{fig:newt_gr_lin}
}
\end{figure}

We can study how the difference between the calculations changes as a function of the
magnitudes of the initial inhomogeneities.  In Fig.~\ref{fig:newt_gr_lin} we
again focus on the density measured by some fiducial observers and show the
fractional difference between the GR and either Newtonian or linear
perturbation results for a range of values for $\bar{\delta}$.  For
$\bar{\delta}\leq10^{-3}$ a correction quadratic in $\bar{\delta}$ to the
density is evident in the GR versus linear comparison, which reaches as high as
tens of percent at the end. However, the difference from the Newtonian results
is roughly an order of magnitude smaller. The $<1\%$ difference from this case
seems to be consistent with being due to numerical truncation error, as
illustrated in the Appendix.  Since $|\psi_N| \lesssim 2\times10^{-3}$ for these cases, even
though the deviations from the background density are large, gravity is still
weak.

\begin{figure}
\begin{center}
\includegraphics[width=\columnwidth,draft=false]{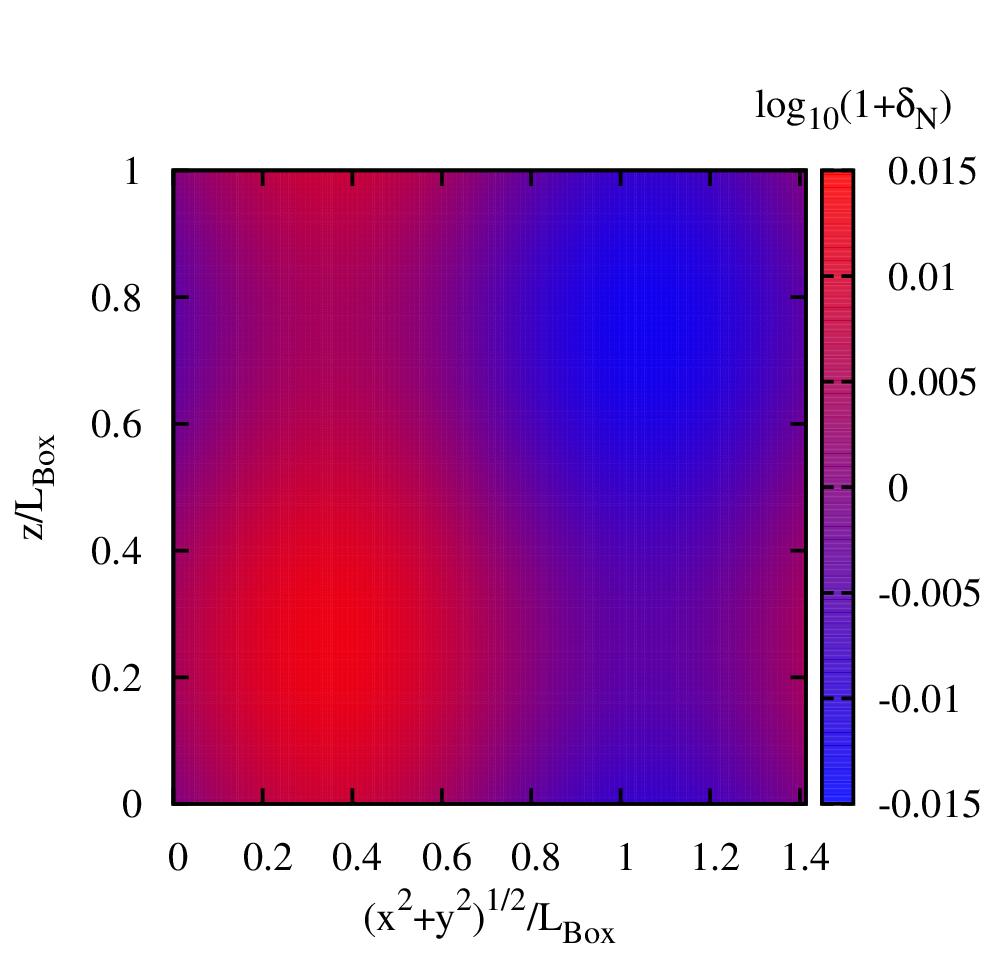}
\includegraphics[width=\columnwidth,draft=false]{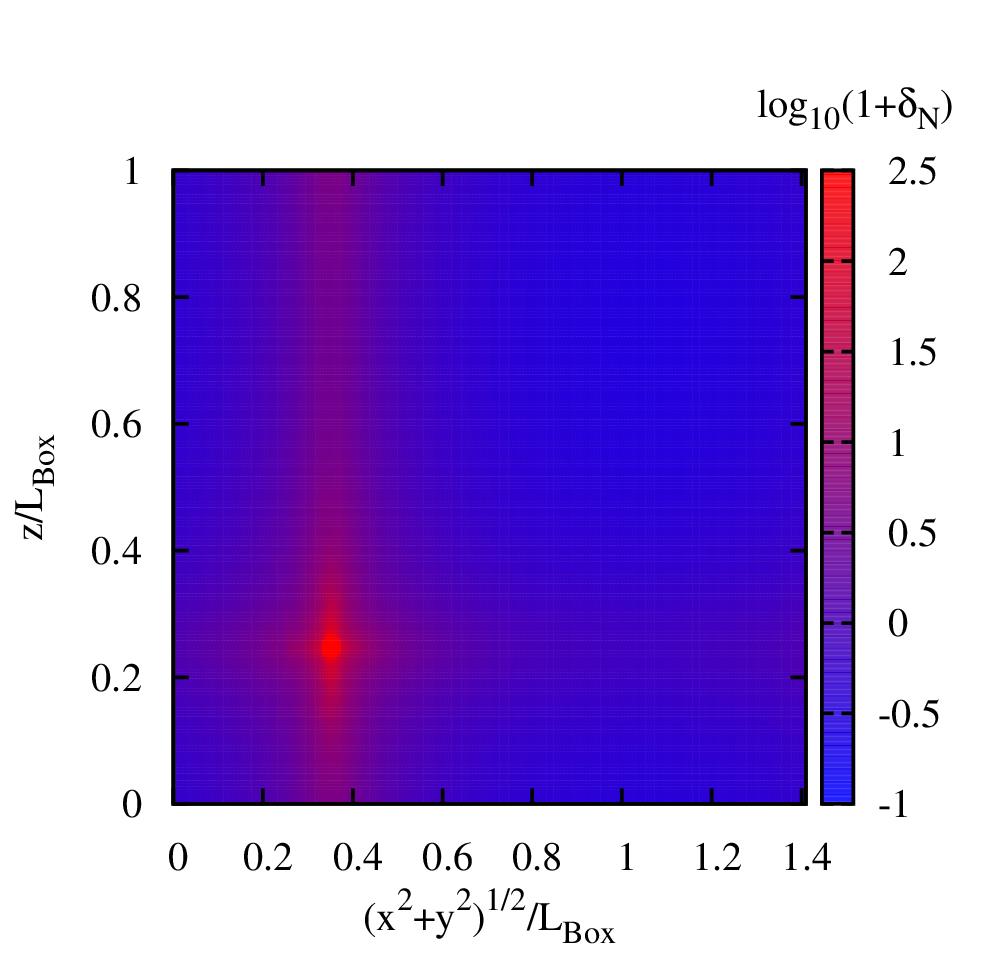}
\end{center}
\caption{
Two-dimensional slice of $\delta_N$ from the Newtonian simulations at the initial ($a=1$; top)
and final ($a=60$; bottom) times from the $\bar{\delta}=10^{-2}$ case.
\label{fig:10m2_density}
}
\end{figure}

The case with $\bar{\delta}=10^{-2}$ is more extreme,
with initial amplitude density perturbations that exceed the equivalent 
values in the standard $\Lambda$CDM model by roughly a factor of a 100. 
Figure \ref{fig:10m2_density} shows the initial 
and final density contrast $\delta_{N}$ from N-body simulations. Though the density at the
point of maximum underdensity, which develops into a void, is again very close in the
Newtonian and GR calculations, strong differences can be seen in the maximum
overdensity at late times, with the Newtonian density exceeding that of the GR.
In fact, in the Newtonian case a massive halo forms around this point with
$\psi_N\sim-0.2$, while in the GR case the fluid density grows without bound,
so the approximation of weak gravity is definitely breaking down. 
The divergence between the GR and Newtonian densities coincides 
quite well with the moment of halo formation predicted by the standard theory of 
spherical collapse, i.e. $a\approx 56$, at which the linearly extrapolated density contrast 
equals $1.686$. Before that, the GR and Newtonian simulations return 
fully consistent densities at all times until $a\approx 50$ when $\delta\approx 10$.

This discrepancy at halo formation is, of course, (at least partially) due to just
the differing treatments of matter.
In the particle case, after shell crossing at $a\approx 59$ at the point of maximum
overdensity, the velocity dispersion goes from zero, to having a value of
$\sigma_v\approx0.2$--$0.3$. 
In the pressureless fluid treatment, there is nothing to halt the collapse,
and we do not continue the calculation beyond this point. 
Thus, for this case we do not compare the GR and Newtonian results past the 
point where multistream regions form.

\begin{figure}
\begin{center}
\includegraphics[width=\columnwidth,draft=false]{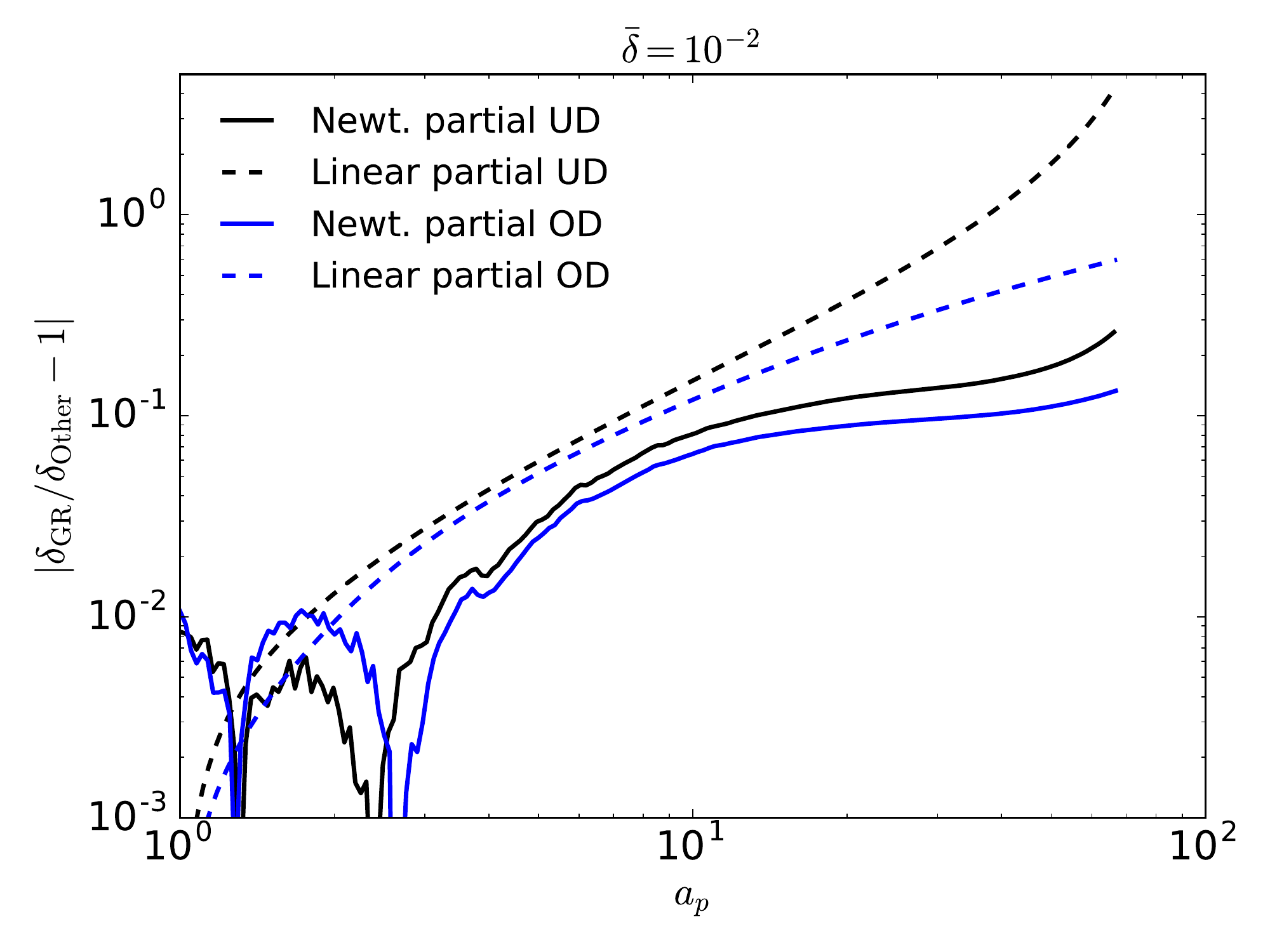}
\end{center}
\caption{
The fractional difference between the density contrast $\delta_{\rm obs}$
measured by an observer comoving with the fluid initially at
$(x,y,z)=(\pi/k)\times(-1/2,1/2,1/2)$ (black lines) and $(\pi/k)\times(-1/2,-1/2,1/2)$
(blue lines) in the GR calculation compared to Newtonian simulations (solid
lines) or linear perturbation theory (dotted lines) for $\bar{\delta}=10^{-2}$. 
\label{fig:10m2_dens_other}
}
\end{figure}

Similar but less extreme
differences can also be found in other overdense and underdense regions in the
$\bar{\delta}=10^{-2}$ case.  As shown in Fig.~\ref{fig:10m2_dens_other},
roughly $10\%$ differences appear at e.g. $(x,y,z)=(\pi/k)\times(-1/2,1/2,1/2)$
and $(\pi/k)\times(-1/2,-1/2,1/2)$ (and similarly at the permutations of the
Cartesian directions).  At both the overdense and underdense points shown,
$|\delta_{\rm obs}|$ is larger in the Newtonian case. 
In contrast to the lower
density cases, these differences do not appear to be due to resolution effects
(though things do begin to become under-resolved at very late times at the
center of the halo; see the Appendix for details).

\begin{figure}
\begin{center}
\includegraphics[width=\columnwidth,draft=false]{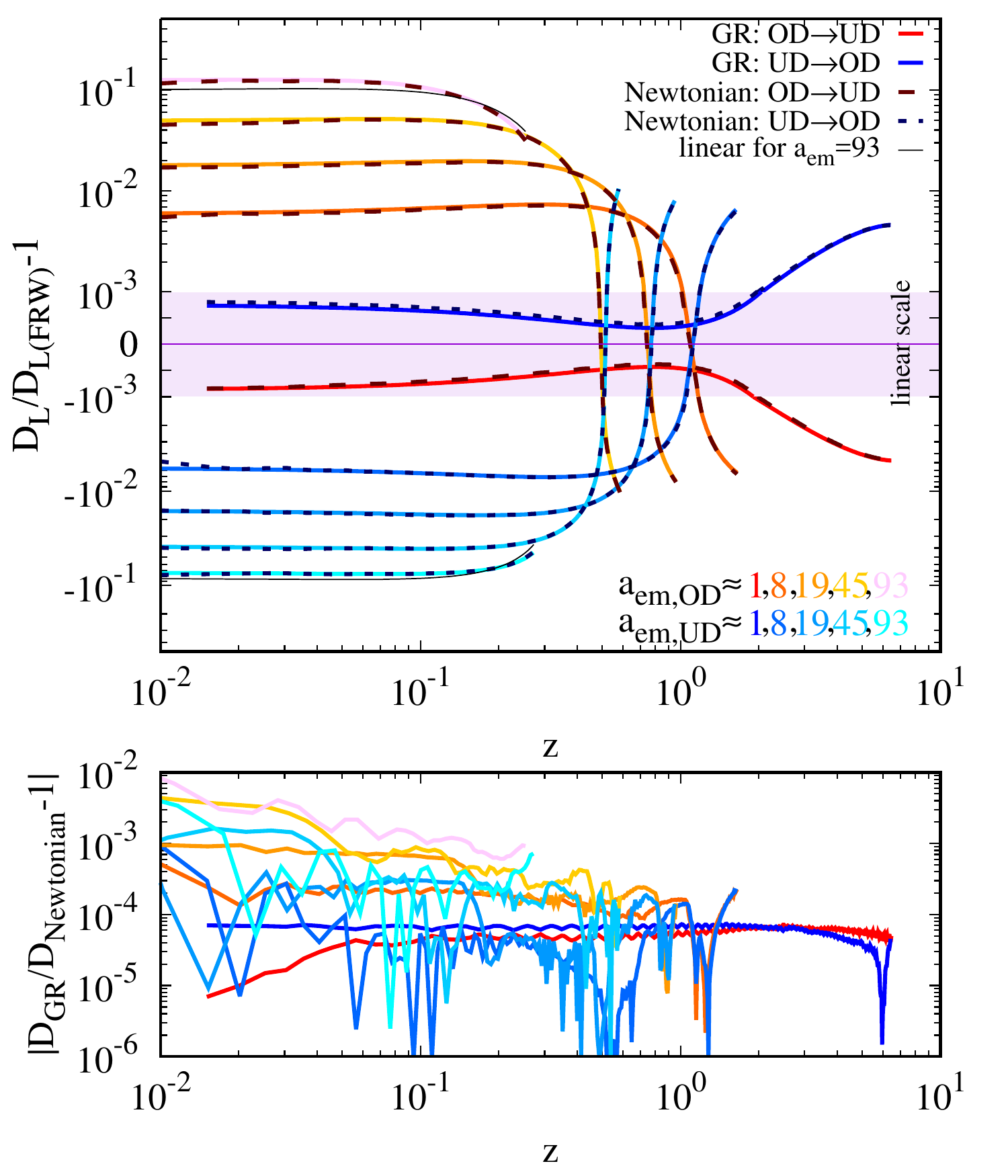}
\end{center}
\caption{The luminosity distance to the global maximum (OD) or minimum (UD) of 
the density field in the model with $\bar{\delta}=10^{-3}$. Distances are computed 
for an ensemble of observers located along the line joining the two critical points. 
Approximate values of the scale factor at subsequent moments of light emission 
are provided in the bottom right corner. The upper panel shows deviation of the 
GR/Newtonian distances from the corresponding observable based on the FRW 
metric and the bottom panel shows the fractional difference between the GR and Newtonian 
calculations. The shaded band indicates a linear scale of the vertical axis. 
The general-relativistic (thick solid lines) and Newtonian (dashed lines) 
simulations return consistent distances with compatible deviations from the FRW 
model and the predictions of linear Newtonian evolution (thin black lines, shown 
only for the latest emission time).
\label{fig:dL_delta10m3}
}
\end{figure}

\begin{figure}
\begin{center}
\includegraphics[width=\columnwidth,draft=false]{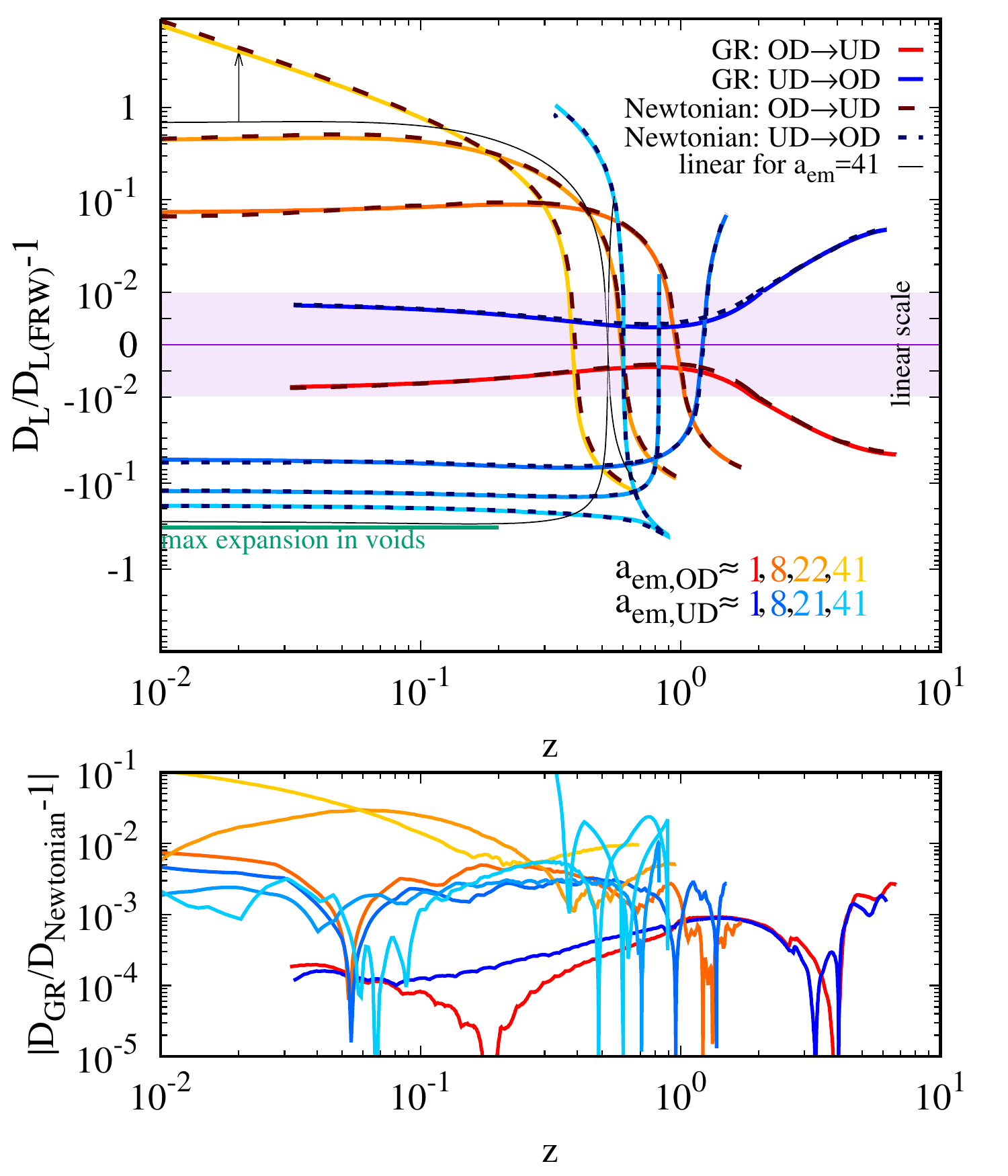}
\end{center}
\caption{The same as Fig.~\ref{fig:dL_delta10m3} but for the case with $\bar{\delta}=10^{-2}$. 
The green line shows the maximum super-Hubble flow in empty voids, i.e. $\delta_{N}\rightarrow -1$, 
based on Newtonian calculations.
\label{fig:dL_delta10m2}
}
\end{figure}

We can also compare the propagation of light as a measure of the differences
between the two cases. To illustrate this, we note that our setup has a line of
symmetry connecting the point of maximum overdensity and underdensity along which null
geodesics will propagate.  Hence, we can consider beams of light rays emitted
by an observer at the overdensity (underdensity) at specified intervals of
proper time and specified frequency, and calculate the redshift and luminosity
distance, as seen by observers comoving with matter, as the beam propagates
and finally reaches the underdensity (overdensity).
This is shown in Figs.~\ref{fig:dL_delta10m3} and \ref{fig:dL_delta10m2} for
initial conditions with $\bar{\delta}=10^{-3}$ and $10^{-2}$, respectively.
In the former we can see that, similar to the density contrast, at later
times, once the perturbations have entered the horizon and begun collapsing, 
there are significant, order $10\%$, deviations from the homogeneous value of $D_L$,
and also noticeable nonlinear corrections.  However, again, the Newtonian and GR
calculations agree quite well, with differences $\lesssim 1\%$, compatible with being
due to numerical error. 
 
For the $\bar{\delta}=10^{-2}$ case shown in Fig.~\ref{fig:dL_delta10m2}, there
are even stronger effects from the inhomogeneities, with order unity deviations
away from the linear perturbation value for $D_L$, and also some cases where
the light rays are blueshifted as they approach the large, collapsing overdensity, causing
$z$ to decrease. Again, the nonlinear Newtonian and GR calculations track each other 
quite well. However, there are noticeable differences which can be likely 
ascribed to the violation of the weak field regime,
inside the halo of the Newtonian simulation
(see the cases demonstrating light propagation inside the overdensity at late times: 
the yellow curve at small redshifts and the light blue curve at high redshifts).

Figure \ref{fig:dL_delta10m2} demonstrates how the nonlinear phase of evolution, both in GR and Newtonian simulations, 
develops asymmetry between light propagation inside the overdensity and underdensity. For the latter, both simulations  
consistently show the emergence of a super-Hubble flow---a linear relation between redshift and distance with 
the effective Hubble constant $H_{\rm eff}>H_{0}$ at $z\lesssim0.2$. Homogeneity of the super-Hubble expansion 
(in contrast to the local Hubble flow at the overdensity) reflects the fact that matter evacuation not only increases the density contrast 
in voids but also homogenizes the residual matter distribution \citep{Woj2016}. Our results demonstrate that both GR 
and Newtonian simulations provide a consistent description of this mechanism. In addition, we can see that the effective 
Hubble constant $H_{\rm eff}$ at subsequent emission times converges to its asymptotic value given by the maximum 
expansion in voids predicted in Newtonian gravity (the green line in Fig.~\ref{fig:dL_delta10m2}), i.e. 
$H_{\rm eff}\rightarrow (3/2)H$ for $\Omega_{\rm m}=1$ \citep[][]{Ber1997}.

\subsection{Initial conditions with range of scales}
\begin{figure}
\begin{center}
\includegraphics[width=\columnwidth,draft=false]{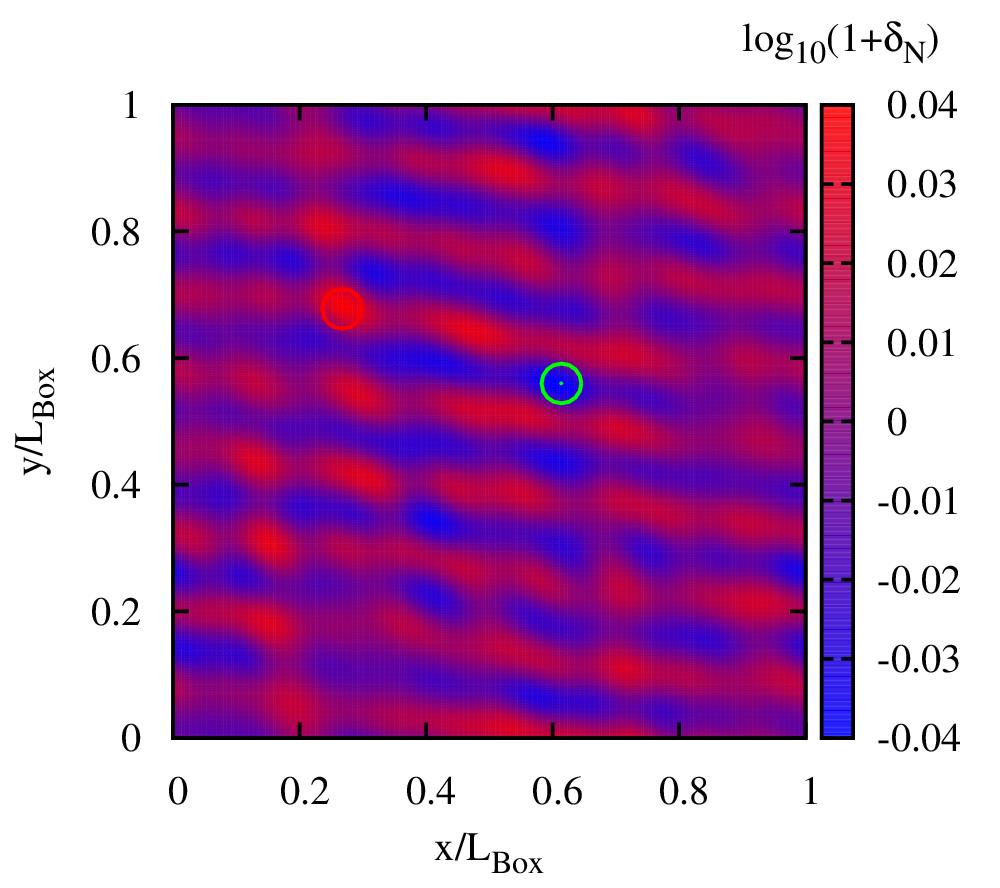}
\includegraphics[width=\columnwidth,draft=false]{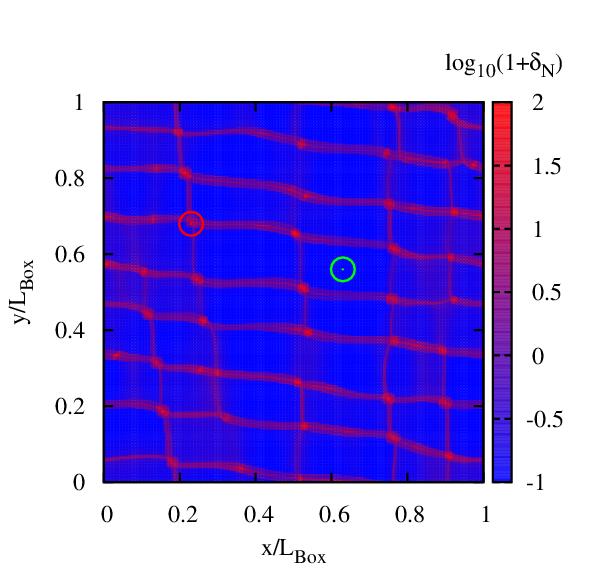}
\end{center}
\caption{
Two-dimensional slice of $\delta_N$ from the Newtonian simulations at the initial ($a=1$; top)
and final ($a=100$; bottom) times from the $\dps=10^{-2}$ case. The red and green points
indicate the positions of fiducial observers initially at points of maximum overdensity and underdensity, 
respectively.
\label{fig:ps_density}
}
\end{figure}

We next consider more general initial conditions that begin with
variations over a range of length scales, to further study possible coupling
between short and long length scales.  In particular, we use the power spectrum
initial conditions described in Sec.~\ref{sec:cases} which have density
variations on wavelengths ranging from 4 times to one-third the initial
Hubble radius at two different amplitudes, which we label $\dps=10^{-3}$ and
$\dps=10^{-2}$.  To illustrate this we show a slice through the initial and
final Newtonian density contrast from the higher amplitude case in
Fig.~\ref{fig:ps_density}. As is evident in the bottom panel, this model generates
a network of halos with $\delta_{N}\gtrsim10^{2}$ and voids with
$\delta_{N}\sim-0.9$.

\begin{figure}
\begin{center}
\includegraphics[width=\columnwidth,draft=false]{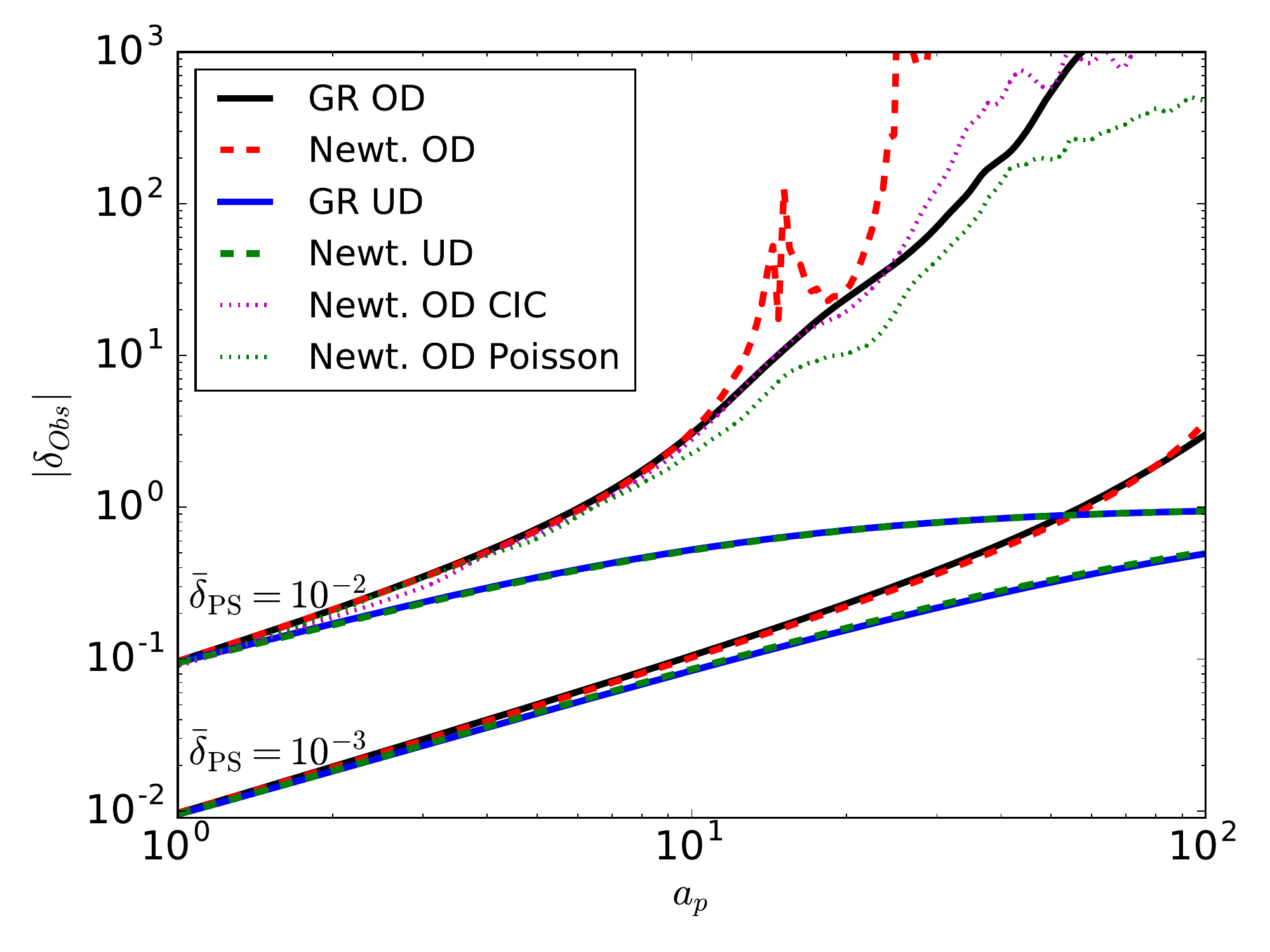}
\end{center}
\caption{
The density contrast as measured by an observer comoving with matter at the
points of maximum overdensity (black and red curves) and underdensity (blue and green
curves) shown in Fig.~\ref{fig:ps_density} for two cases with
$\dps=10^{-3}$ and $\dps=10^{-2}$.  
The general-relativistic (solid lines) and Newtonian (dashed lines)
calculations closely track each other except at the overdensity
in the largest amplitude case where a noisy multistream region forms. 
In such regions the N-body density contrast is sensitive to the particular
density estimator used, which we illustrate by also including the cloud-in-cell
density estimate, as well as the one obtained from the potential through the 
Poisson equation (dotted lines), for one case.
\label{fig:ps_delta}
}
\end{figure}

As an indication of the evolution of these cases, in Fig.~\ref{fig:ps_delta} we
show the density (relative to a FRW solution) seen by fiducial observers
comoving with matter, at the initial points of minimum and maximum density
(marked in Fig.~\ref{fig:ps_density}). As in the previous cases, there is broad
agreement between the Newtonian and GR results even as the inhomogeneities
become large. At the point of maximum density, the velocity dispersion becomes
nonzero in the particle case at $a\approx15$ and eventually reaches a value of 
$\sigma_v\sim0.04$. 
As nonlinear structure and multistream regions (in the 
N-body case) form, the density value at the overdensity becomes noisy, as well
as fairly sensitive to resolution and the density estimator used. 
We illustrate
this latter point in Fig.~\ref{fig:ps_delta} by also including the density
estimate for one of the N-body cases using two alternative methods: the
cloud-in-cell method~\cite{Springel2005}, and by calculating the density from the
gravitational potential through the Poisson equation. Because of this, in what follows we will
concentrate on comparing the propagation of light rays in the
respective spacetimes.

\begin{figure*}
\begin{center}
\includegraphics[width=2.1\columnwidth,draft=false]{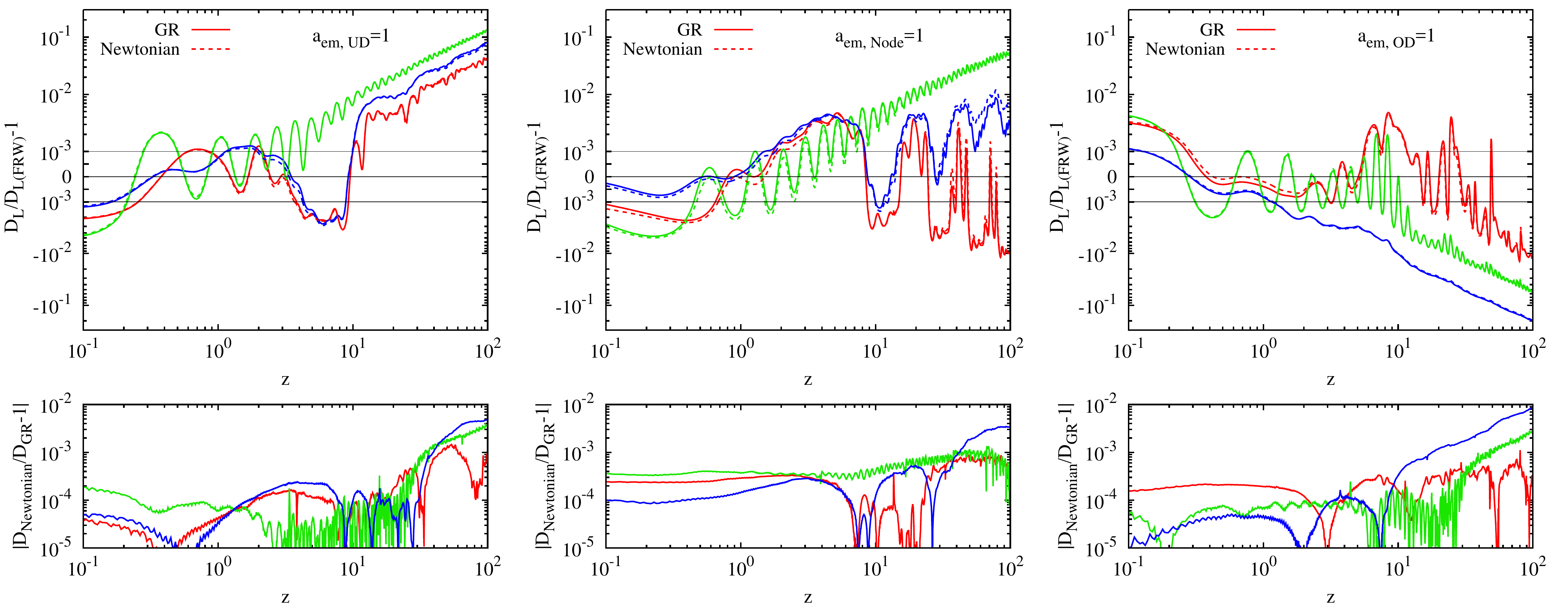}
\end{center}
\caption{The luminosity distance to the global minimum (left panels), 
the node point (middle panels) and the global maximum (right panels) of 
the initial density field in the cosmological model with $\dps=10^{-3}$. 
Distances are computed for observers located along photon rays emitted 
at the scale factor $a_{\rm em}=1$ in the $x$, $y$ and $z$ directions. The 
upper panels show deviations between the GR and Newtonian distances from 
the corresponding observables based on the unperturbed FRW metric, 
and the bottom panels show the fractional difference between the GR and 
Newtonian results. Both the GR and Newtonian simulations return 
consistent distances with nearly the same deviations from the FRW 
model.
\label{fig:dL_PS10m3}
}
\end{figure*}

\begin{figure*}
\begin{center}
\includegraphics[width=2.1\columnwidth,draft=false]{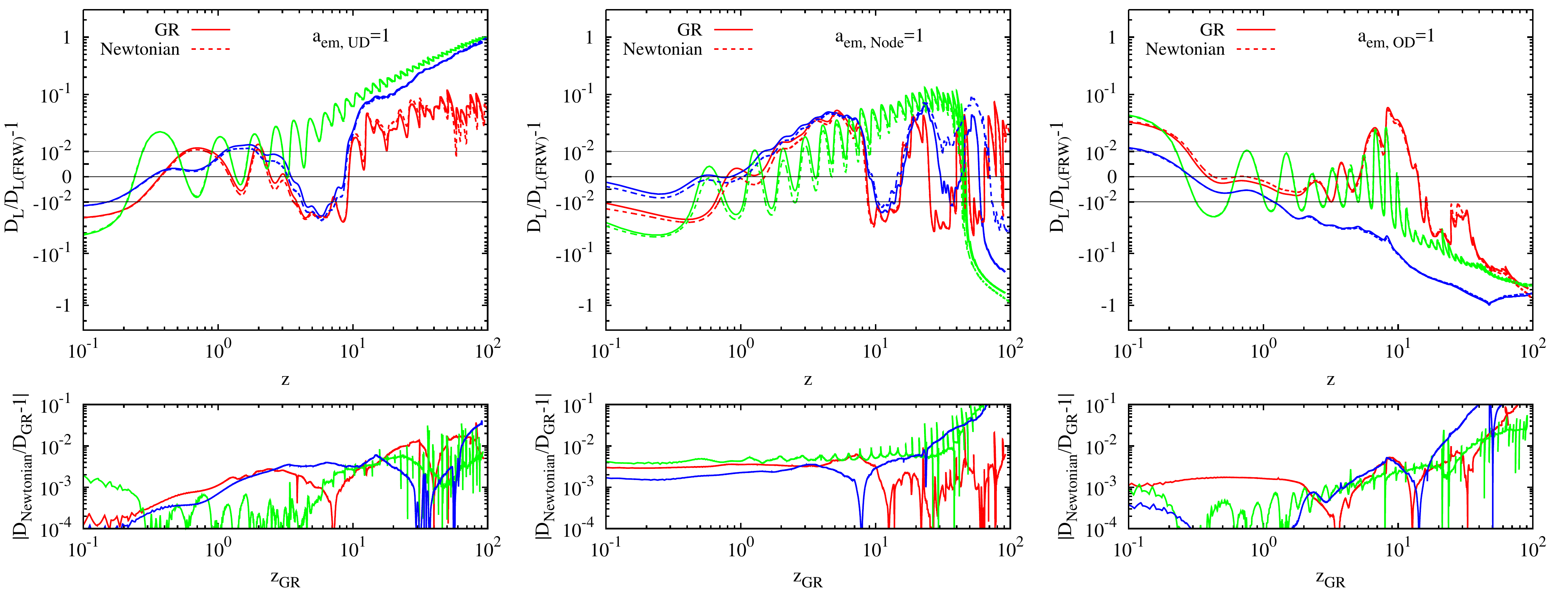}
\end{center}
\caption{The same as Fig.~\ref{fig:dL_PS10m3} but for the case with $\dps=10^{-2}$. 
The fractional differences are computed at the same value of affine parameters and plotted 
as a function of redshift from the GR simulations. The corresponding fractional differences 
in redshift are smaller than $10^{-2}$ at all times.
\label{fig:dL_PS10m2}
}
\end{figure*}

Since it does not have the discrete symmetry of the initial conditions
considered in Sec.~\ref{subsec:simple_id}, for this setup we consider a set of
light rays with initial positions at the points of minimum and maximum
density, as well as an intermediate point with $\delta_N=0$. For 
each position we consider light rays with initial
velocities pointing in plus and minus each of the $x$, $y$, and $z$ coordinate
directions that then propagate throughout the simulations.  We show the
luminosity-redshift values---again, as measured by observers comoving with
matter---for a representative set of these in Fig.~\ref{fig:dL_PS10m3}
for $\dps=10^{-3}$ and Fig.~\ref{fig:dL_PS10m2} for $\dps=10^{-3}$.  The effect
of short and longer wavelength inhomogeneities is evident in higher and lower
frequency components of the deviations from the homogeneous values of $D_L$
versus $z$.  In the higher amplitude case (Fig.~\ref{fig:dL_PS10m2}) strongly
nonlinear effects are apparent, including instances of decreasing redshift with
increasing luminosity distance and ``lensing" which causes $D_L$ to pass
through zero.  However, these features are captured by both the full GR
spacetime and the one reconstructed from the Newtonian solution. Though there
are some evident quantitative differences between the values of $D_L$ obtained
in the two cases, these can be primarily attributed to numerical truncation
error due to the small-scale structure eventually not being well resolved.
This is illustrated in the Appendix (in particular, Fig.~\ref{fig:ps_dlz}),
where we include lower resolution results.

Figures \ref{fig:dL_PS10m3} and \ref{fig:dL_PS10m2} clearly demonstrate a difference between light 
propagating primarily in voids (left panels) or overdense regions (right panels). The cumulative effect 
of tidal forces (corresponding to the Riemann tensor in the GR calculations) makes the photon rays diverge in 
the former case or converge in the latter. This in turn manifests itself as demagnification (increased distances) 
or magnification (decreased distances), respectively. Our results show that this cumulative lensing effect 
is consistently described both in fully GR computations and in (relativistic) ray tracing on the effective
spacetime of the Newtonian simulations.

\section{Discussion and Conclusion}
\label{sec:conclusion}

We have systematically compared cosmological models of structure formation calculated 
using the full Einstein equations, to those using Newtonian gravity on a homogeneously 
expanding background. We considered a suite of globally flat cold dark matter models (Einstein--de Sitter models) 
with a range of density perturbations on scales comparable to the Hubble horizon at the initial 
time. Starting with consistent initial conditions based on the correspondence between 
GR and Newtonian cosmology in the linear regime of the density evolution, we evolved the models to a highly 
nonlinear phase using both numerical GR coupled to hydrodynamics and N-body techniques. The GR and Newtonian 
simulations were then compared in terms of the density field and the properties of light propagation. 
The former was consistently calculated for an ensemble of freely falling observers located at various points 
of the initial density field. The latter was quantified by solving the geodesic equations describing bundles 
of light rays emitted from a set of sources, consistently defined in both simulations. Every bundle of 
geodesics was then used to determine cosmological distance as a function of redshift, as measured 
by free-falling observers located along the photon path.

Our comparison between GR and Newtonian simulations in the highly nonlinear
phase does not reveal any significant differences, as long as the Newtonian
potential does not violate the weak field assumption.  Our resolution studies
show that in most cases any apparent differences between the GR densities and
their counterparts from Newtonian simulations---typically sub-percent in the level of the inhomogeneities---are due to truncation error in 
the simulations. In general, the fractional differences between the two
decrease in higher resolution runs. The only exception is one case with high
density regions with shell crossings.
In this one case, we were not able to continue the GR fluid calculation past
the time where shell crossing occurs in the Newtonian N-body calculation.  This
is predominantly due to the lack of full conformity between the treatment of
matter in the hydrodynamical and particle description.  A more thorough
comparison of GR and Newtonian simulations into this regime will probably require
using particles (or hydrodynamics) with both treatments of gravity.  In the
other cases considered here, particle versus fluid differences due to multistream
regions were subdominant to numerical truncation error.

Despite some noticeable differences between GR and Newtonian in regions 
where the weak gravity assumption is violated, we do not see any dissimilarities 
between gravitational
collapse in the GR and Newtonian frameworks. In particular, in the model with the
highest amplitude of the initial density field ($\bar{\delta}=10^{-2}$), the
Newtonian evolution closely resembles the GR collapse until $\delta_{\rm N}\sim
10$. Taking the moment of abrupt growth of density as the halo formation time
(the first shell crossing in Newtonian simulations), we demonstrated that both
GR and Newtonian simulations point to the halo formation time that is
consistent with the standard spherical collapse model.  This is in contrast to
\citep{Bentivegna:2015flc} which considered a similar setup, and reported a lag between gravitational collapse
in GR and the standard (Newtonian) spherical model. 
Our study also suggests that the results
of~\cite{Giblin:2015vwq,Giblin:2016mjp} are similarly in a regime where the
observed nonlinear effects should be well captured by a nonlinear
Newtonian calculation. 

In our study we have made use of the ``abridged dictionary"
of~\cite{Chisari:2011iq,Green:2011wc} which relates the quantities from  a
Newtonian cosmology calculation to the general-relativistic spacetime metric
and stress-energy tensor that they should approximate.  Though this
correspondence is only strictly applicable at the linear level, as argued
in~\cite{Green:2011wc}, the corrections should be small even with large
inhomogeneities, as long as they occur on small scales and the gravitational
potential remains small.  Our study demonstrates, for the first time, by means
of explicit comparison of fully GR and Newtonian cosmological simulations, that
indeed this is the case, even beginning with inhomogeneities on scales comparable
to the Hubble horizon and continuing to the highly nonlinear regime of the
density evolution.  The Newtonian simulations are able to arrive at these
solutions with considerably less computational expense, both because of the
fewer number and less complicated nature of evolution equations and
because roughly 100 times fewer time steps have to be taken.  In most
cases, we found the differences between the Newtonian and relativistic
calculations to be dominated simply by numerical errors. Though here we focused
on somewhat simplified setups with a limited range of length scales, and hence
less stringent resolution requirements, production-level structure-formation
N-body simulations typically have numerical errors that are comparable or
worse~\cite{Heitmann:2007hr,Schneider:2015yka}, meaning it will be quite
challenging to make such errors subdominant to any relativistic effects. 
Having said that, we emphasize that we have focused on somewhat simplified
setups in this work, and our study does not exhaust all possible initial
conditions, nor probe the effects of other types of matter or cosmological
parameters such as dark energy, curvature, etc. Other, more relativistic types
of matter, e.g. neutrinos, may exhibit stronger differences.

The close resemblance between our GR and Newtonian simulations is especially
prominent in the comparison distances calculated from ray tracing (in both cases,
based on solving geodesic equations describing bundles of light rays).  Both
simulations consistently capture all effects, giving rise to noticeable
deviations from the observables based on the FRW metric including the enhanced
(suppressed) expansion in overdense (underdense) regions (see
Figs.~\ref{fig:dL_delta10m3} and~\ref{fig:dL_delta10m2}) and the
demagnification (magnification) in voids (overdensities) (see
Figs.~\ref{fig:dL_PS10m3} and~\ref{fig:dL_PS10m2}). Although numerical errors
appear to be larger for some cases featuring particularly strong lensing, ray
tracing yields remarkably similar characterization of these lensing events in
both simulations.

Most observables used in cosmological inference are not based on directly
integrating the geodesic equation but are rather derived under a number of
simplifying assumptions, e.g. linearity of density evolution and the Born
approximation of thin lenses adopted commonly in lensing calculations.
However, it is not obvious whether forgoing such approximations when
calculating observables can introduce significant corrections or not, and this
is something currently under investigation
(e.g.~\cite{Petri:2016qya,Fabbian:2017wfp}). Several recent studies have
attempted to address this problem by combining fully GR cosmological
simulations with full ray tracing \citep{Giblin:2016mjp,Gib2017lens}. Our
results suggest, however, that any possible corrections to the standard
cosmological observables may stem from inaccurate ray tracing adopted in the
standard framework rather than from a genuine difference between GR and
Newtonian evolution of the density field.  Therefore, to test the standard
framework for calculating cosmological observables, it may be worth further
exploring the easier and computationally less expensive strategy of using
standard N-body simulations to reconstruct a spacetime and directly
integrating geodesics on it, as we do here
(see~\cite{Koksbang:2015ima,Koksbang:2015jba} for work along these lines).  The
same strategy can also be useful in theoretical considerations involving
cosmological models with large-scale perturbations exceeding the limits imposed
by the standard $\Lambda$CDM model.  For example, our study shows that models
with large-scale local voids can feature substantially higher, locally measured,
Hubble constants and thus are able to reproduce basic properties of recently
studied cosmological models with an observationally constrained relation between
the cosmological redshift and cosmic scale factor, dubbed redshift remapping
\citep{Woj2016rr,Woj2017rr}.

\acknowledgments
We thank Stephen Green, Matt Johnson, Luis Lehner, and Jim Mertens for stimulating
discussions.  Simulations were run on the Sherlock Cluster at Stanford
University and the Bullet Cluster at SLAC.  This research was supported in part
by Perimeter Institute for Theoretical Physics.  Research at Perimeter
Institute is supported by the Government of Canada through the Department of
Innovation, Science and Economic Development Canada and by the Province of
Ontario through the Ministry of Research, Innovation and Science. R.W. 
was supported by a grant from VILLUM FONDEN (Project No. 16599).

\appendix
\section*{Appendix: Numerical error results}

\begin{figure}
\begin{center}
\includegraphics[width=\columnwidth,draft=false]{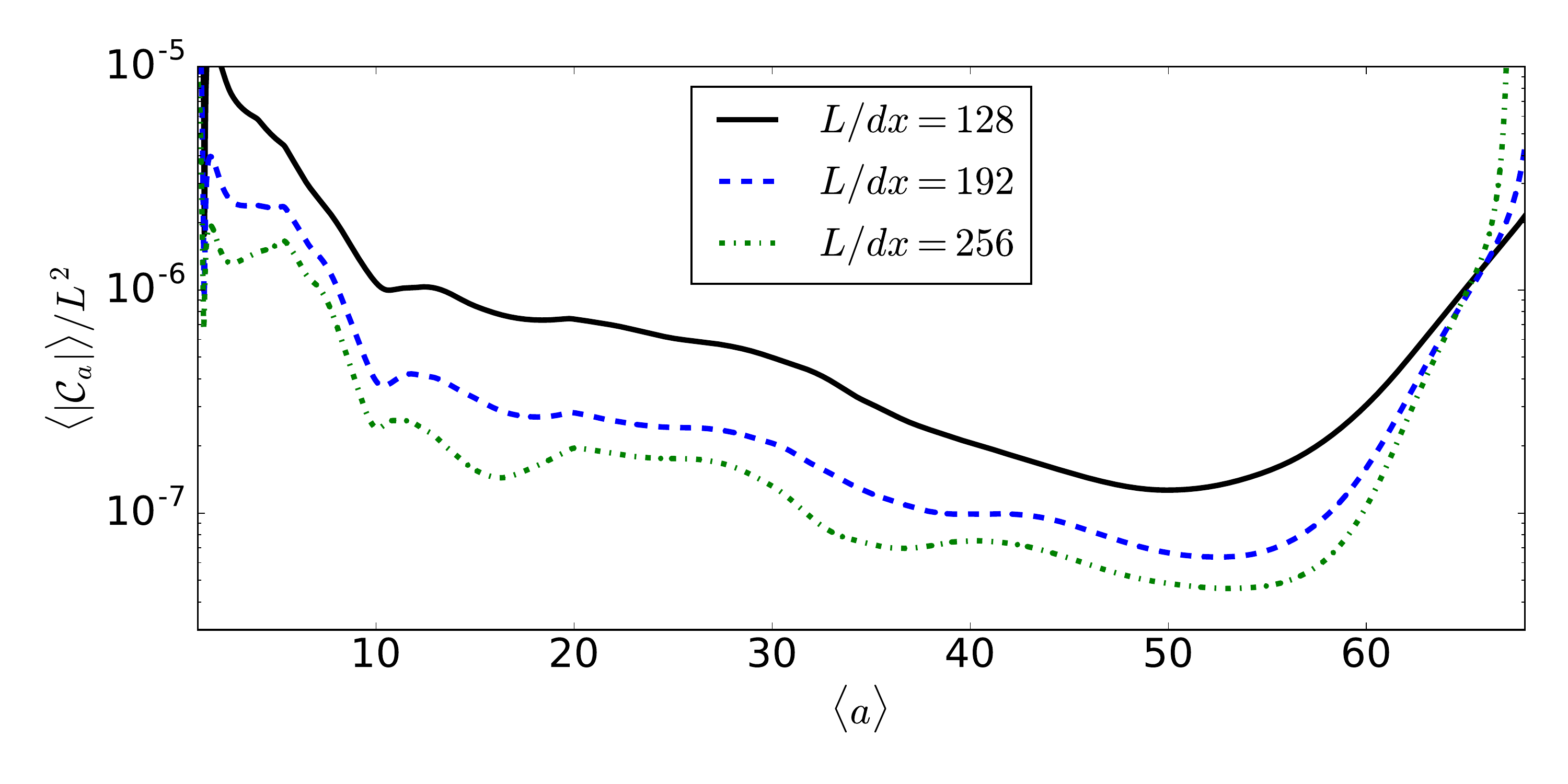}
\includegraphics[width=\columnwidth,draft=false]{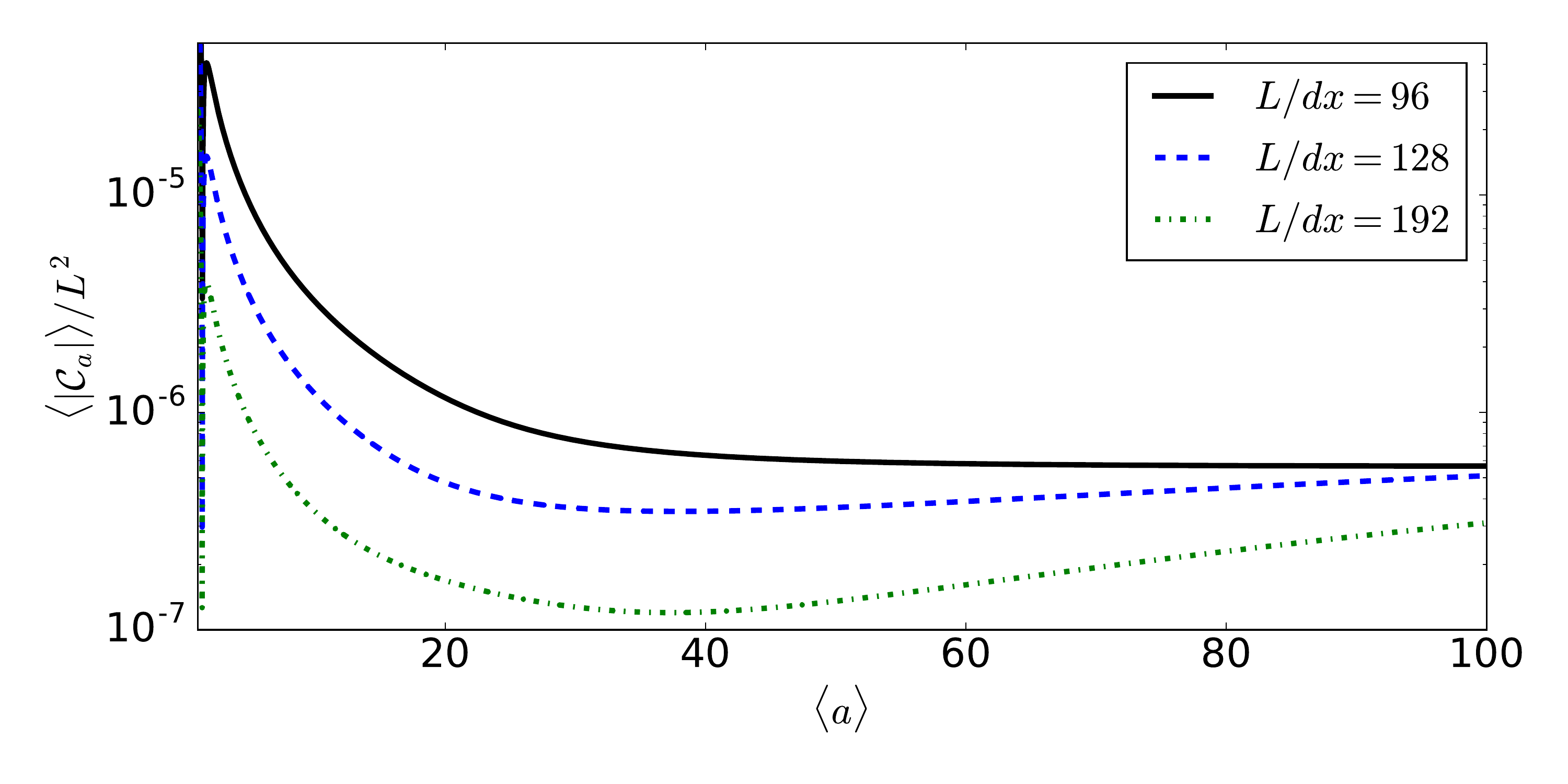}
\end{center}
\caption{
The volume average of the norm of the constraint violation $\mathcal{C}_a:=\Box
x_a - H_a$ as a function time (parametrized by a volume-averaged measure of the
effective scale factor), for simulations with $\bar{\delta}=10^{-2}$ (top) and
$\dps=10^{-3}$ (bottom) at three different resolutions.  The decrease in
constraint violation with increasing resolution is consistent with roughly
second-order convergence.
\label{fig:cnst_conv}
}
\end{figure}

In this appendix we give some details on numerical convergence and error
estimates.  It is important to determine if the differences seen between the
various quantities compared between the Newtonian and GR simulations are due to
the differences in the underlying equations or just to differences in the
numerical truncation error.  In order to estimate this, we run selected cases at
multiple resolutions.  For the Newtonian N-body simulations we use a low,
medium, and high resolution with $64^3$, $128^3$, and $256^3$ particles,
respectively. For most of the GR calculations we
use a low, medium, and high resolution with a grid with $96^3$, $128^3$, and
$192^3$ cells, respectively.  For the cases with the largest amplitude
inhomogeneities (the $\bar{\delta}=10^{-2}$ and $\dps=10^{-2}$ cases) we use
$128^3$, $192^3$, and $256^3$ grid cells.  To illustrate convergence, in
Fig.~\ref{fig:cnst_conv} we show the magnitude of the generalized harmonic
constraint violation for several cases.  The convergence of this quantity to
zero with increasing resolution is a nontrivial check that the constraint
equations at the initial time, and the evolution equations, are being solved
with sufficient resolution (see~\cite{Pretorius:2004jg}).

\begin{figure*}
\begin{center}
\includegraphics[width=0.66\columnwidth,draft=false]{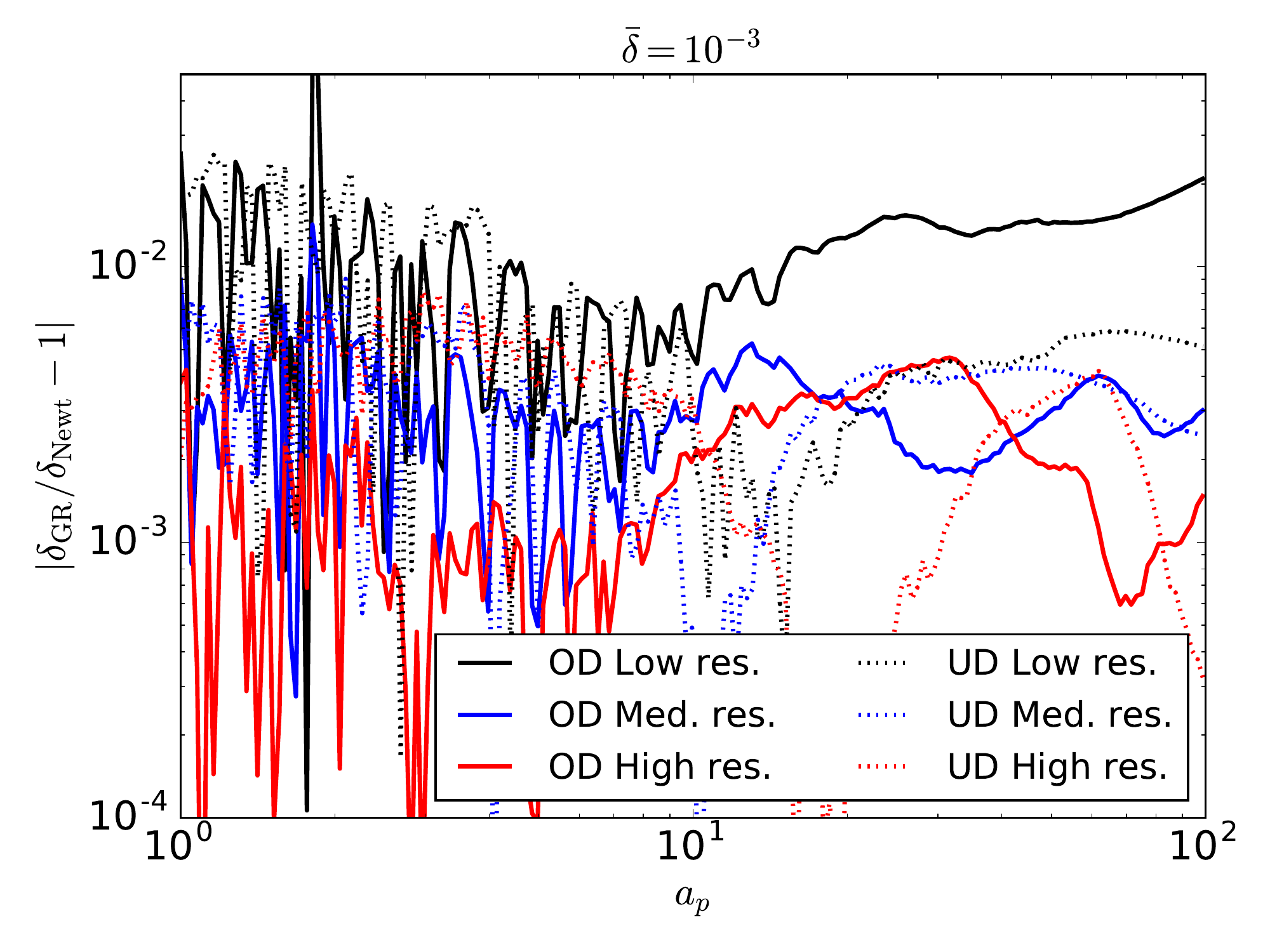}
\includegraphics[width=0.66\columnwidth,draft=false]{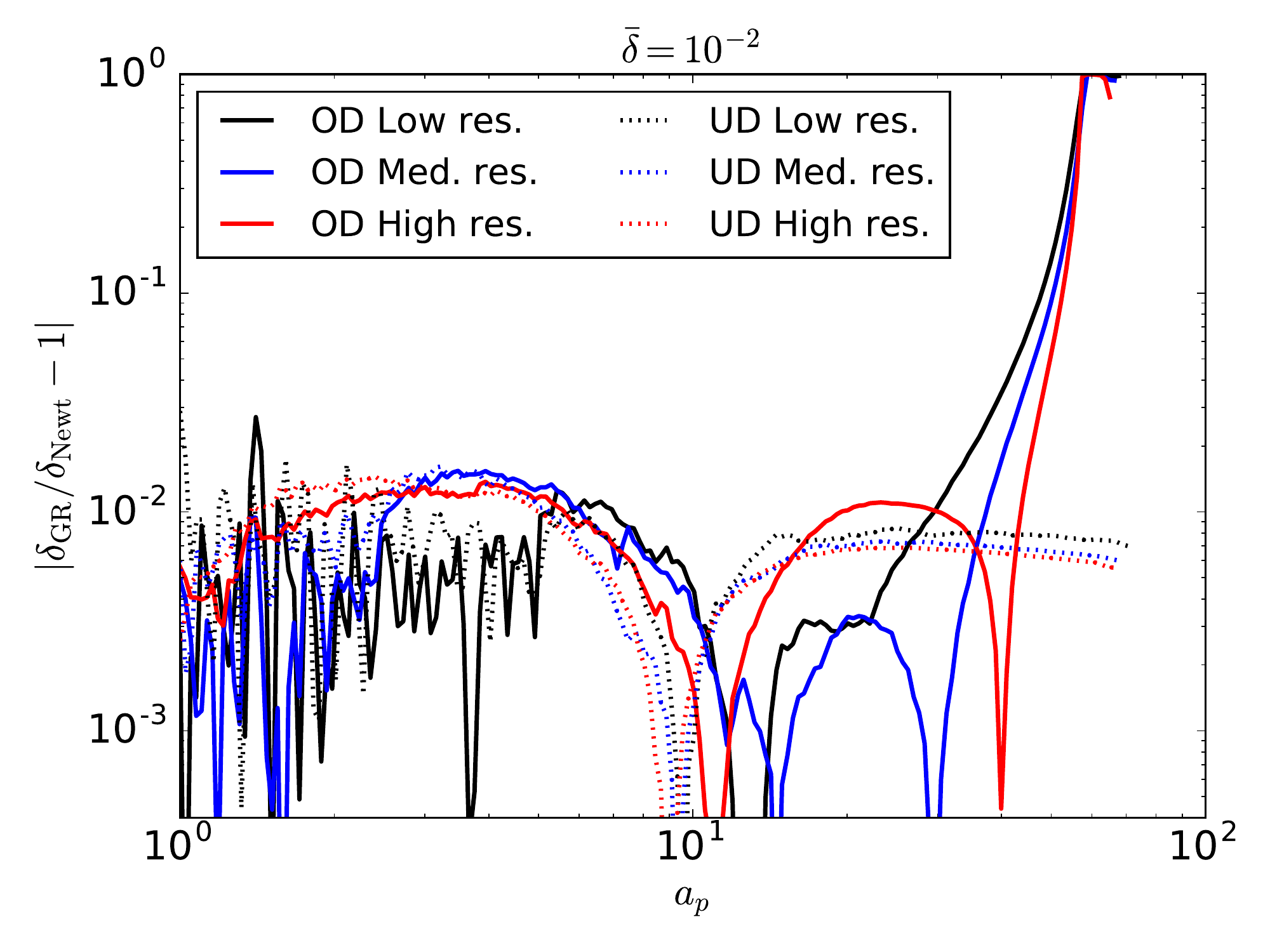}
\includegraphics[width=0.66\columnwidth,draft=false]{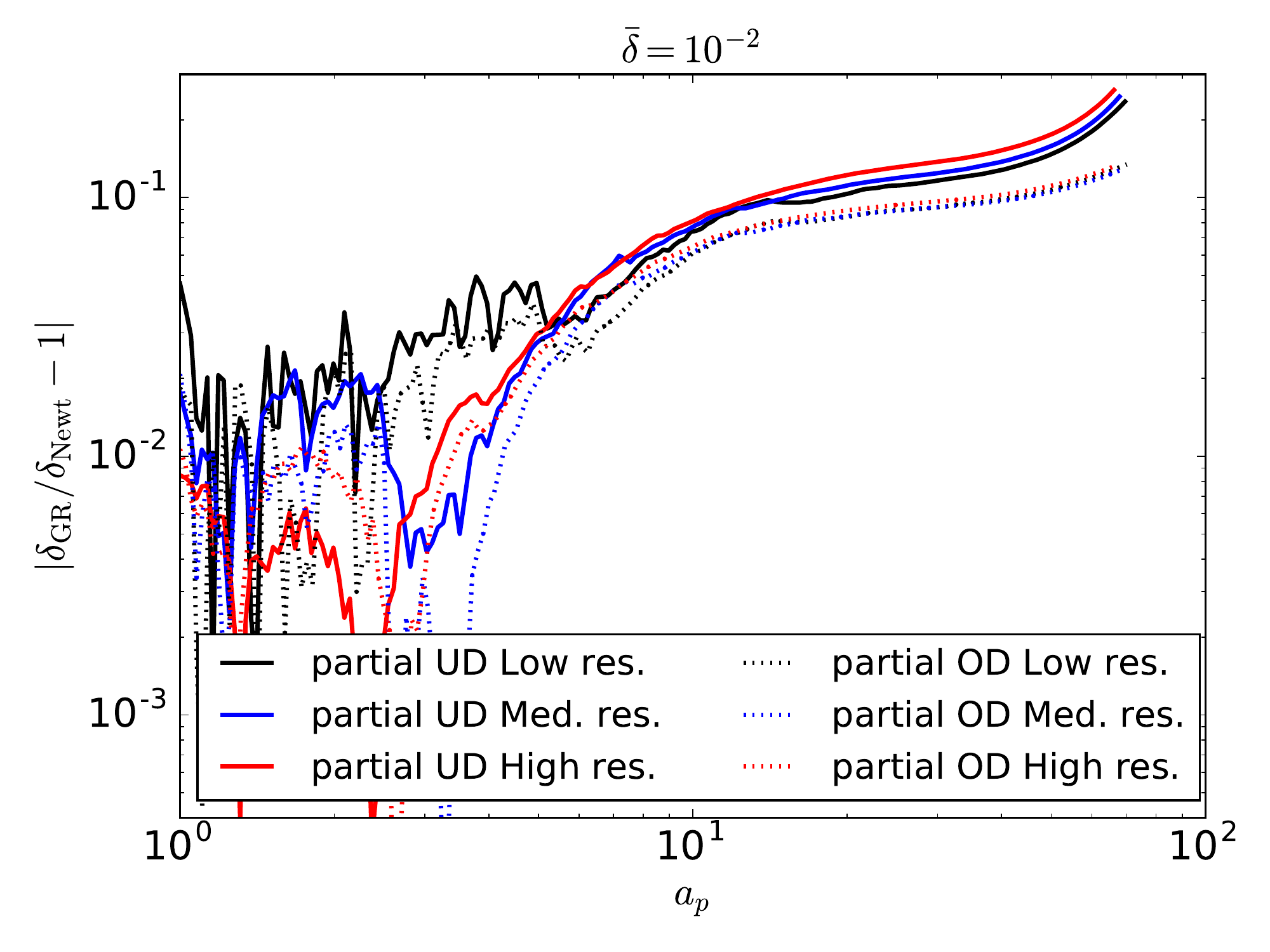}
\end{center}
\caption{ The relative difference in density contrast $\delta_{\rm obs}$
between the GR and Newtonian simulations, at three different sets of
resolutions. The left and middle panels show the points of maximum overdensity and
underdensity for $\bar{\delta}=10^{-3}$ and $\bar{\delta}=10^{-2}$,
respectively (similar to Fig.~\ref{fig:newt_gr_lin}), while the right panel
shows the same intermediate points from the $\bar{\delta}=10^{-2}$ case as
in Fig.~\ref{fig:10m2_dens_other}.
\label{fig:delta_res_diff}
}
\end{figure*}

The Newtonian and GR calculations will have different truncation error, with
different dependence on resolution.  However, to give a rough estimate, we show
the difference between several quantities in the Newtonian and GR simulations at
multiple resolutions.  In Fig.~\ref{fig:delta_res_diff} we show the difference
in the density contrast measured at the overdensity and underdensity for
$\bar{\delta}=10^{-3}$ and $\bar{\delta}=10^{-2}$ (left and middle panels; cf.
Fig.~\ref{fig:newt_gr_lin}), as well as some intermediate points for
$\bar{\delta}=10^{-2}$ (right panel; cf. Fig.~\ref{fig:10m2_dens_other}).
For many of the cases, the difference between the two calculations decreases as
the resolution of the respective simulations is increased, indicating that the
discrepancy is primarily attributable to truncation error.  However, at late
times in several of the $\bar{\delta}=10^{-2}$ cases, differences in the
density that are consistent with increasing resolution are apparent. 

\begin{figure}
\begin{center}
\includegraphics[width=\columnwidth,draft=false]{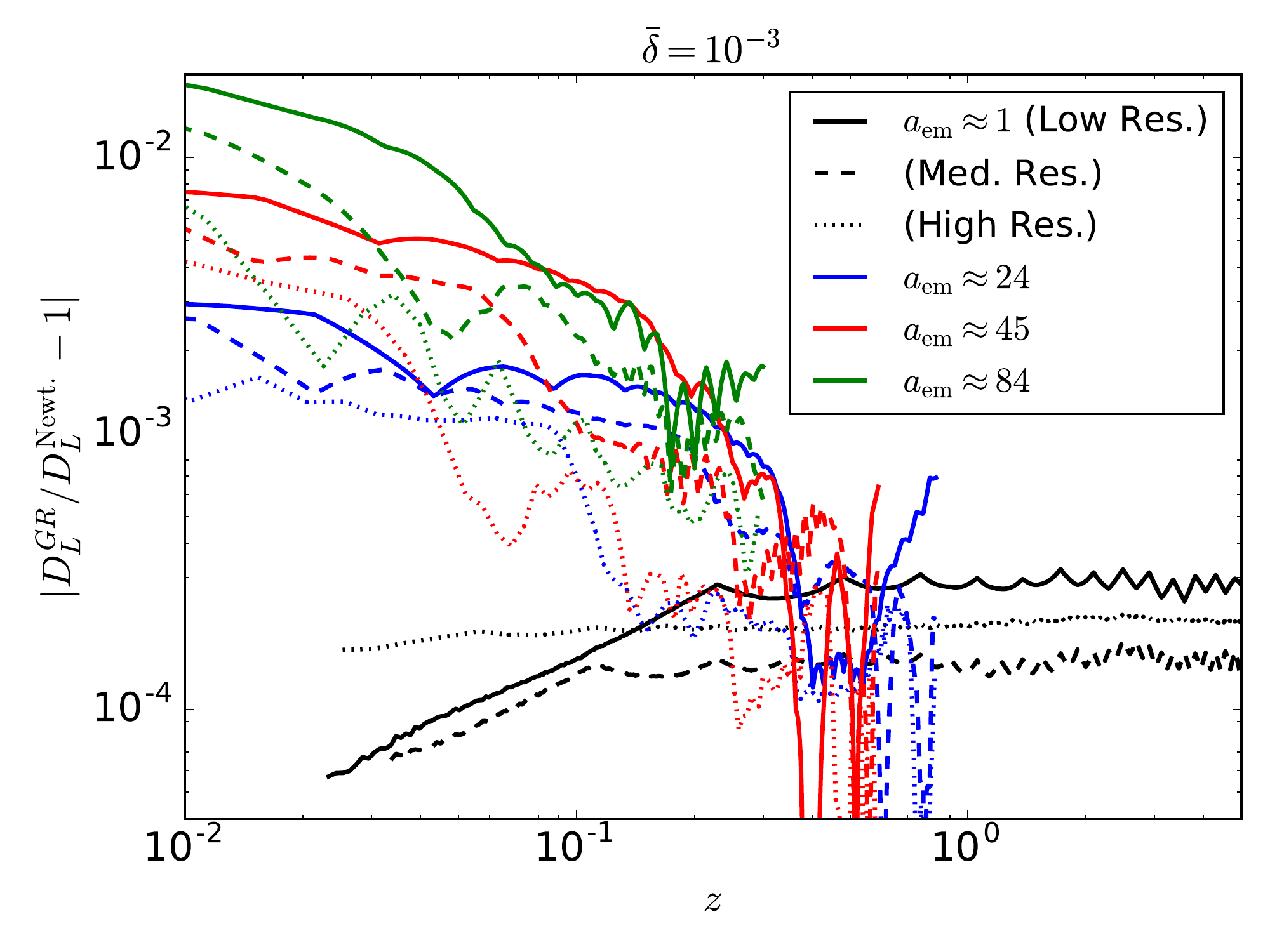}
\includegraphics[width=\columnwidth,draft=false]{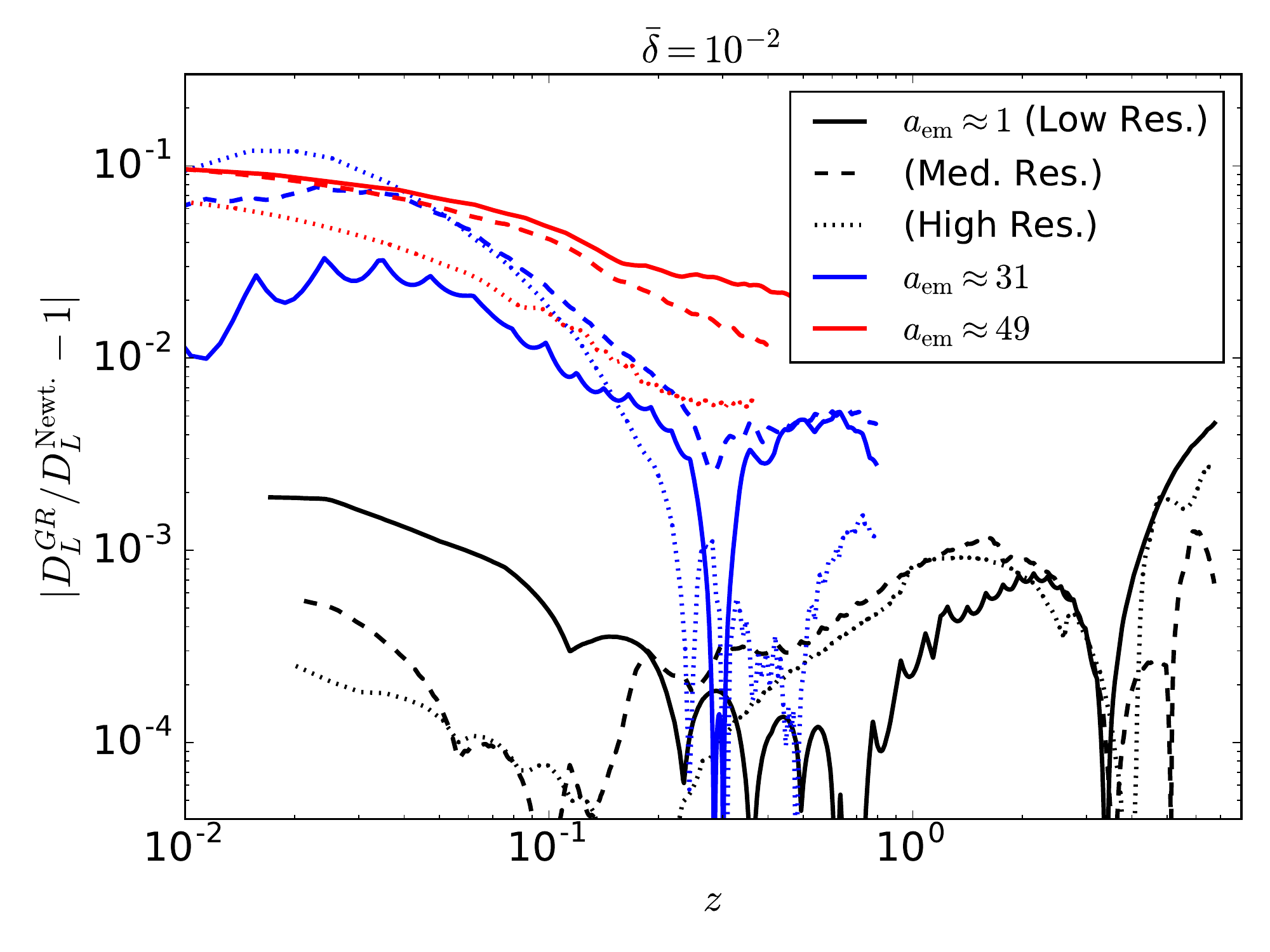}
\end{center}
\caption{The relative difference in the luminosity versus redshift between
the GR and Newtonian simulations at different resolutions for light rays
emitted at the point of maximum overdensity that propagate to the point of
maximum underdensity for the $\bar{\delta}=10^{-3}$ (top) and
$\bar{\delta}=10^{-2}$ (bottom) cases, similar to
Figs.~\ref{fig:dL_delta10m3} and \ref{fig:dL_delta10m2}.
\label{fig:delta_dlz_res_diff}
}
\end{figure}

We also show the dependency of the redshift-luminosity relations on resolution
in Fig.~\ref{fig:delta_dlz_res_diff}. Again, for the $\bar{\delta}=10^{-3}$ case
shown in the top panel, the difference between the GR and Newtonian results
decreases noticeably with increasing resolution, indicating that the $\lesssim
1\%$ differences seen in Fig.~\ref{fig:dL_delta10m3} are likely dominated by
truncation error. Here we just show the light rays beginning at the overdensity
and ending at the underdensity, but the reverse ones are similar.  However, for the
$\bar{\delta}=10^{-2}$ case shown in the bottom panel, there are some significant
differences in the luminosity distance in the vicinity of the overdensity as it
collapses at later times, though the differences diminish as the light rays
propagate farther away.  

\begin{figure}
\begin{center}
\includegraphics[width=\columnwidth,draft=false]{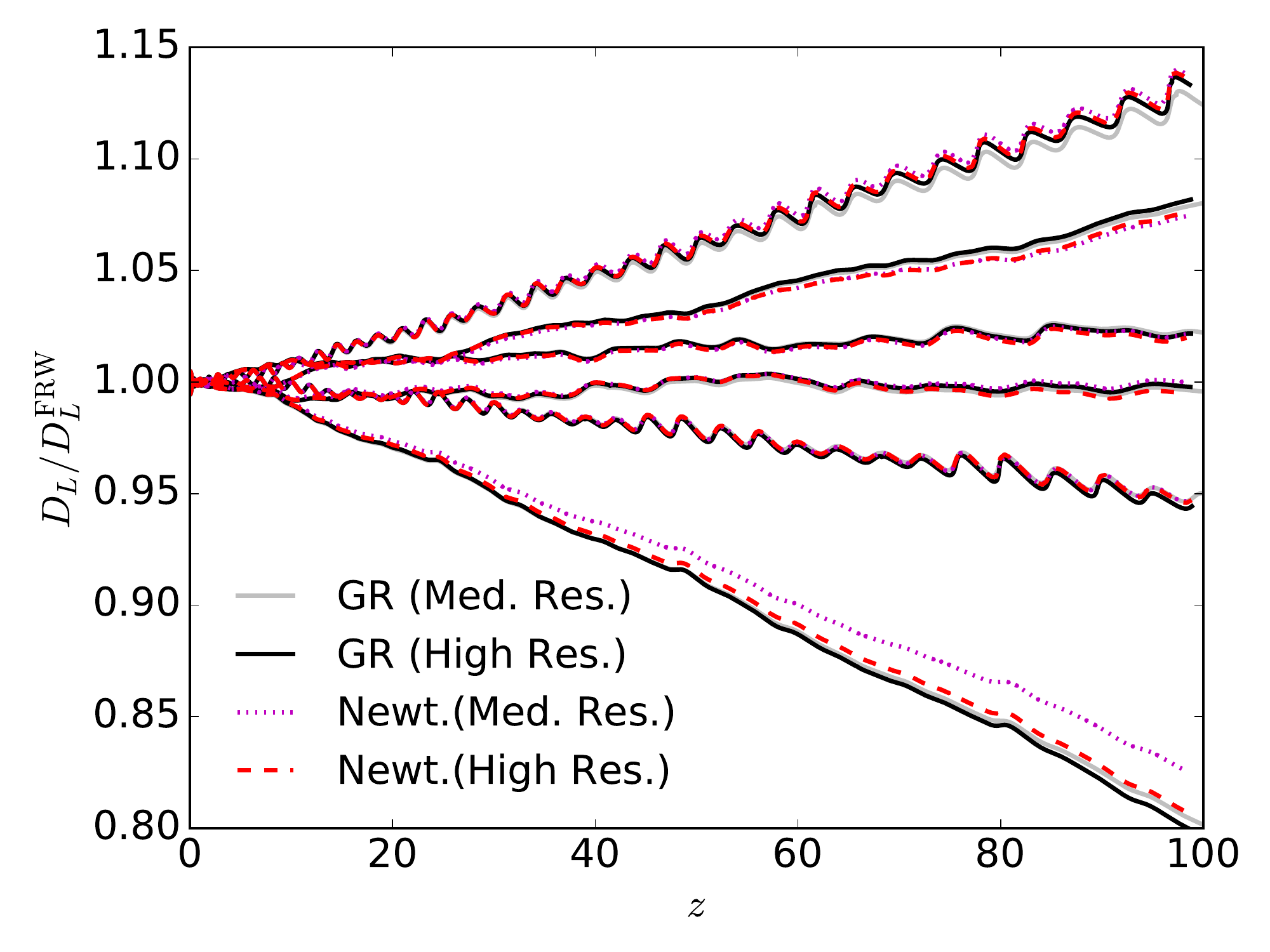}
\includegraphics[width=\columnwidth,draft=false]{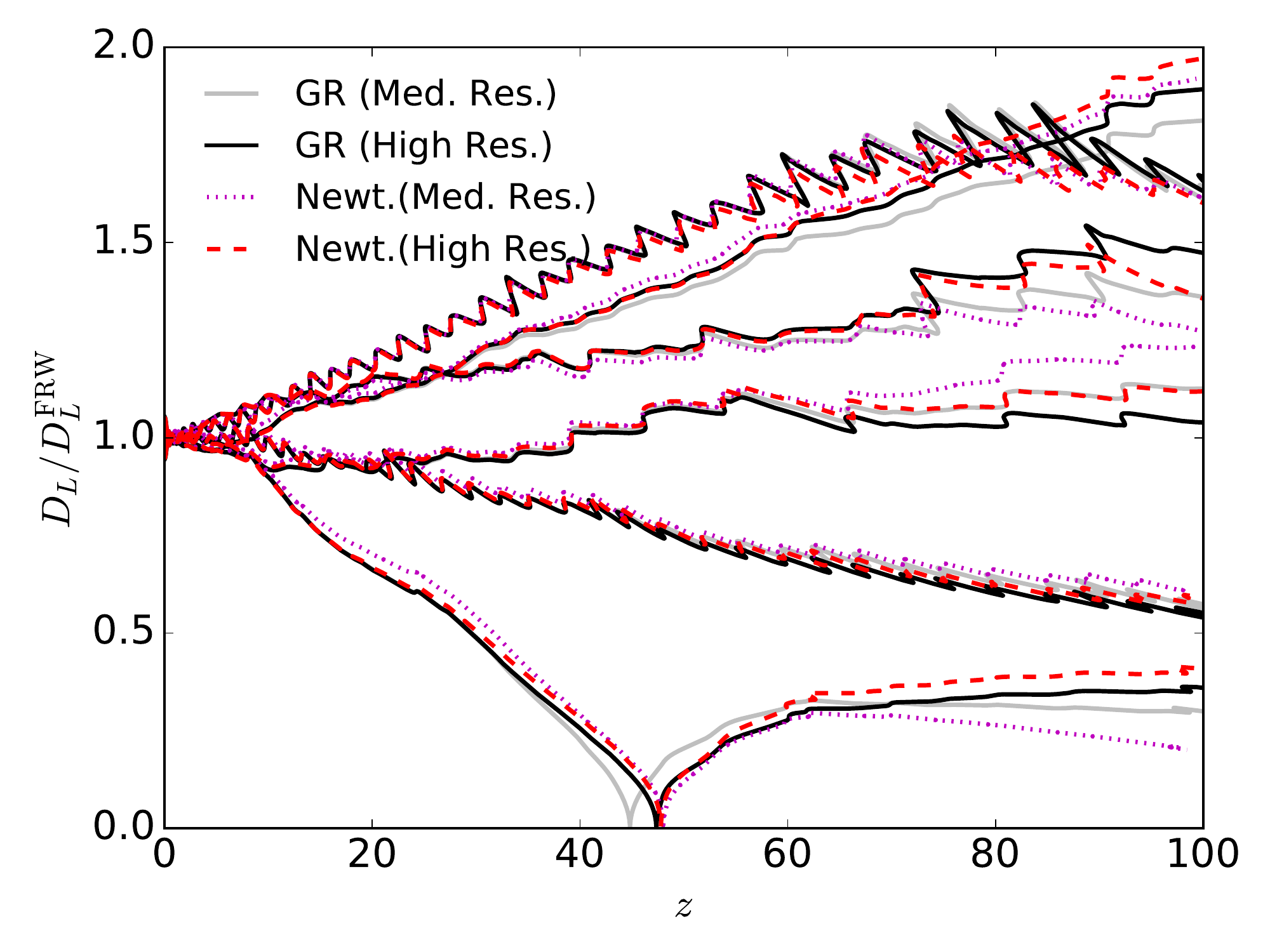}
\end{center}
\caption{
The luminosity distance (normalized by the FRW value) versus redshift for a
representative set of null rays in the $\dps=10^{-3}$ (top) and 
$\dps=10^{-2}$ (bottom) cases. We show results from both the highest resolution
GR (solid black lines) and Newtonian (dashed red lines) simulations, as well as
lower resolution results from both cases (grey solid and magenta dashed lines,
for GR and Newtonian, respectively) to indicate the magnitude of the numerical
truncation error. 
\label{fig:ps_dlz}
}
\end{figure}

Finally, as an indication of the magnitude of truncation error in the
power-spectrum initial conditions simulations, in Fig.~\ref{fig:ps_dlz} we show
high- and medium-resolution results for redshift versus luminosity for $\dps=10^{-3}$
and $\dps=10^{-2}$. Here it can be seen that the difference between the GR and
Newtonian values is both comparable to the difference with resolution and
diminishes as the resolution is increased. This is true both for $\dps=10^{-3}$
(top panel), where the differences are small, and $\dps=10^{-2}$ (bottom panel),
where stronger resolution-dependent effects are evident at late times.

\bibliographystyle{apsrev4-1}
\bibliography{ref}

\end{document}